\documentclass[final,5p,times,twocolumn,authoryear]{elsarticle}

\usepackage{hyperref}
\usepackage{graphicx}
\usepackage{deluxetable}

\usepackage{amssymb}

\usepackage[labelfont=bf,justification=raggedright,singlelinecheck=false]{caption}
\journal{Earth and Planetary Science Letters}

\begin{document}

\begin{frontmatter}

\title{High-temperature miscibility of iron and rock during terrestrial planet formation}

\author[mymainaddress]{Sean M.Wahl\corref{mycorrespondingauthor} }
\cortext[mymainaddress,mycorrespondingauthor]{Corresponding author}
\ead{swahl@berkeley.edu}

\author[mymainaddress,mysecondaryaddress]{Burkhard Militzer}

\address[mymainaddress]{Department of Earth and Planetary Science, University of California
Berkeley, United States}
\address[mysecondaryaddress]{Department of Astronomy, University of California
Berkeley, United States }

\begin{abstract}
The accretion of a terrestrial body and differentiation of its silicate/oxide mantle from
iron core provide abundant energy for heating its interior to temperatures much higher than
the present day Earth.  The consequences of differentiation on the structure and
composition of planets are typically addressed considering only the interaction of molten
iron with an immiscible `rocky' phase. We demonstrate that mixing in a representative
system of liquid or solid MgO and liquid iron to a single homogeneous liquid occurs at
sufficiently low temperature to be present in the aftermath of a giant impact. Applying
the thermodynamic integration technique to density functional theory molecular dynamics
simulations, we determine the solvus closure temperature for the Fe-MgO system for
pressures up to 400 GPa. Solvus closure occurs at $\sim$4000 K at low pressure, and has
a weak positive pressure dependence, such that its gradient with respect to depth is less
steep than an adiabatic temperature profile. This predicts a new mode of core-mantle
differentiation following the most energetic giant impacts, with exsolution of
iron from the mixture beginning in the outer layers of the planet. We demonstrate that
high-temperature equilibration results in delivery of nominally insoluble Mg-rich
material to the early core. Since MgO is the least soluble major mantle component in iron at
low temperatures, these results may represent an upper bound on temperature for mixing
in terrestrial planets.
\end{abstract}

\begin{keyword}
 planet formation \sep composition \sep differentiation \sep miscibility \sep first
 principles \sep iron \sep MgO \sep solvus closure
\end{keyword}

\end{frontmatter}


\section{Introduction} \label{sec:introduction}

Terrestrial planets are, to first order, made up of a metallic iron core and a mantle
composed of silicate and oxide minerals. Chondritic meteorites show that these materials
initially condensed together from the protoplanetary nebula, but became free to separate
and gravitationally stratify as planetesimals grew.  Numerous scenarios have been put
forward to describe how these reservoirs interact depending on the pressure, extent of
melting, and the specific assumptions of rocky phases
\citep{Stevenson1990,Solomatov2007,Rubie2007}. These typically assume the major components
occur in two immiscible phases. Additionally, most studies assume that element
partitioning between the two phases is similar to that observed in experiments performed
at much lower temperatures \citep{McDonough1995}. In the case of a hot early history of a
growing planet, neither assumption is necessarily correct. At sufficiently high
temperatures, entropic effects dominate and any mixture of materials will form a single,
homogeneous phase. It is therefore necessary to consider a high temperature mixture of
the `rocky' and metallic terrestrial components. The presence of such a mixed phase will
affect the chemistry of iron-silicate differentiation on the early Earth.

Here we consider a simple representative material for the mixed rock-metal phase as a mixture
of Fe and MgO formed via the reaction
$\rm{MgO}_{\rm sol/liq} + \rm{Fe}_{\rm liq} \Rightarrow \rm{FeMgO}_{\rm liq}$. 
We determine the stability of these phases using first-principles calculations.  At a
given pressure, a system with two separate phases can be described in terms of a
miscibility gap. At low temperatures, homogeneous mixtures with intermediate compositions
are thermodynamically unstable and a heterogeneous mixture of two phases with
compositions near the endmembers is preferred. The exsolution gap is bounded by a solvus
that marks the temperature above which a single mixed phase is stable, and the maximum
temperature along the solvus is referred to as the solvus closure temperature. Here we calculate
the Gibbs free energy of the mixture and the endmembers to determine the solvus closure
temperature for mixture similar to the bulk-composition of a terrestrial planet. These
results can inform future work, by providing the conditions where rock-metal miscibility
plays a role in the differentiation of terrestrial planet interiors.

An order of magnitude calculation shows the gravitational energy released in the
formation of an Earth-mass body, if delivered instantaneously, is sufficient to raise
temperatures inside the body by $\sim$40,000 K.  Redistribution of mass within the body
during core formation can account for another $\sim$4,000 K increase. This energy is
released over the timescale of accretion, $\sim$$10^{8}$ years \citep{Chambers1998}, with
efficient surface heat loss through a liquid-atmosphere interface
\citep{Abe1997,Elkins-Tanton2012}. However, simulations of the final stages of planet
growth \citep{Cameron1991,Chambers1998,Canup2000} suggest that near-instantaneous release
of large quantities of energy through giant impacts is  the rule rather than the
exception.  Simulations of the `canonical' moon-forming impact hypothesis
\citep{Canup2004}, in which a Mars-sized body collides with the protoearth, find fractions
of the target's interior shocked well above 10,000 K. More recently, dissipation of
angular momentum from the Earth-Moon system by the evection resonance has loosened
physical constraints on the impact, suggesting that the formation of the moon is better
explained by an even more energetic event than the `canonical' one \citep{Canup2012,Cuk2012}.
It is, therefore, difficult to precisely constrain the temperature of the Earth's
interior in the aftermath of the moon forming impact, much less that of other terrestrial
planets with even more uncertain impact histories. Regardless, there is evidence for
giant impacts throughout the inner solar system, implying temperatures significantly
higher than the present day Earth may have been common. In addition to high temperatures,
giant impacts involve significant physical mixing of iron and rocky materials
\citep{Dahl2010}.  Thus, miscibility may be important even if the impacting bodies have
iron cores that differentiated at lower temperatures and pressures.

Differentiation and core formation is a key event or series of events in terrestrial
planet evolution. The timing and conditions of differentiation have important
consequences for the evolution of the planet, through its effect on the distribution of
elements throughout the planet's interior. The distribution of elements affects the
gravitational stability of solid layers in the mantle, the location of radioactive heat
sources, and the nature of the source of buoyancy driving core convection and magnetic
field generation. Each of these subsequently affect the thermal evolution of the planet.
If this process occurs near the solvus closure temperature, there are likely to be
physical and chemical differences from the processes at conditions where the phases are
completely immiscible. We include a discussion of some of these processes in Section
\ref{sec:discussion}.

\section{Theory and methodology} \label{sec:theory}

Modern high pressure experimental techniques, using static or dynamic compression
techniques, can reach megabar pressures \citep{Boehler2000}. However, experiments at
simultaneous high pressures and temperatures have limitations. Interpretation of mixing
processes during shock wave experiments is difficult, and the samples are not
recoverable. Meanwhile, laser heated diamond anvil cells experience extreme temperature
gradients and require survival of quenched texture to interpret. In both cases, the
methods only cover a small fraction of the $P$-$T$ range expected in the aftermath of a
giant impact.  As a result, simulations based on first-principles theories are an
appropriate means of constraining material properties over a range of such extreme
conditions. 

\subsection{Simulated System}
We performed density functional theory molecular dynamics (DFT-MD) simulations for phases
in a model reaction between liquid iron, and solid (B1) or liquid magnesium oxide. The
change in Gibbs free energy of this system per formula unit FeMgO is described as
\begin{equation} \label{eqn:gibbs}
  \Delta G_{\rm{\rm mix}}  =  \frac{1}{24}G_{\rm{(FeMgO)_{24}}} 
- \frac{1}{32} \left[  G_{\rm{(MgO)_{32}}}  + G_{\rm{Fe_{32}}} \right]
\end{equation}
where $G_{\rm{(MgO)_{32}}}$ and   $G_{\rm{Fe_{32}}}$ are the Gibbs free energies of a
pure MgO and iron endmembers with subscripts referring to the number of atoms in the
periodic simulation cell. $G_{\rm{(FeMgO)_{24}}}$ is the Gibbs free energy of 1:1
stoichiometric liquid solution of the two endmembers. Comparing Gibbs free energies among
a range of compositions, we find the temperature for mixing of the phases in the
1:1 ratio to be a good approximation for the solvus closure temperature.

MgO is the simplest mantle phase to simulate, and a reasonable starting point for a study
of the miscibility of terrestrial materials. Up to $\sim$400 GPa, MgO remains in the
cubic B1 (NaCl) structure \citep{Boates2013}, meaning simulations of only one solid phase
were necessary for the rocky endmember. In order to describe a similar reaction for ${\rm
MgSiO_3}$ perovskite, the Gibbs free energy of MgO and ${\rm SiO_2}$ must also be
calculated to address the possibility of incongruent dissolution of the solid phase
\citep{Boates2013}. More importantly, high pressure experiments observing reactions of
silicates with iron have demonstrated the MgO component has by far the lowest solubility
in iron up to $\sim$3000 K \citep{Knittle1991,Ozawa2008a}. This suggests our results  can
be interpreted as an upper bound for the solvus closure temperature with more realistic
compositions. 

It is worth emphasizing that the mixed FeMgO phase is unlike any commonly studied rocky
phase, in that it does not have a balanced oxide formula. This is by design and is
necessary for the mixing of arbitrary volumes of the metallic and oxide phases.  This is
a separate process from the reaction of the FeO component which transfers O to the
metallic phase at lower temperature, and which is primarily controlled by oxygen fugacity
rather than temperature \citep{Tsuno2013}. We treat the mixed phase as a liquid at all
conditions.  Although we cannot absolutely rule out the possibility of a stable solid
with intermediate composition, such a phase would require a lower Gibbs free energy, and
therefore is consistent with treating our results as an upper bound on the solvus closure
temperature.

\subsection{Computation of Gibbs Free Energies}
The Gibbs free energy of a material includes a contribution from entropy of the system.
Since entropy is not determined in the standard DFT-MD formalism, we adopt a two step
thermodynamic integration method, used in previous studies
\citep{Wilson2010,Wilson2012a,Wahl2013,Gonzalez2014}.  The thermodynamic integration technique
considers the change in Helmholtz free energy for a transformation between two systems
with governing potentials $U_a\left(\mathbf{r_i}\right)$ and
$U_b\left(\mathbf{r_i}\right)$. We define a hybrid potential
$U_{\lambda}=\left(1-\lambda\right)U_a+\lambda U_b$, where $\lambda$ is the fraction of
the potential $U_b\left(\mathbf{r_i}\right)$. The difference in Helmholtz free energy is
then given by
\begin{equation} \label{eqn:td_int}
  \Delta F_{a\to b}\equiv F_b - F_a = \int_{0}^{1}{d\lambda\,\langle U_b\left(\mathbf{r_i}\right) -
  U_a\left(\mathbf{r_i}\right) \rangle_{\lambda}}
\end{equation}
where the bracketed expression represents the ensemble-average over configurations,
$\mathbf{r_i}$, generated in simulations with the hybrid potential at constant volume and
temperature. This technique allows for direct comparisons of the Helmholtz free energy of
DFT phases, $F_{\rm DFT}$, by finding their differences from reference systems with a known
analytic expression, $F_{\rm an}$. 

In practice, it is more computationally efficient to perform the calculation $\Delta
F_{\rm an\to DFT}$ in two steps, each involving an integral of the form of Eqn.
\ref{eqn:td_int}. We introduce an intermediate system governed by classical pair
potentials, $U_{\rm cl}$, found by fitting forces to the DFT trajectories
\citep{Wilson2010,Izvekov2004}. For each pair of elements, find the average force in bins
of radial separation and fit the shape of a potential using a cubic spline function. We
constrain the potential to smoothly approach zero at large radii and use a linear
extrapolation at small radii, where the molecular dynamics simulations provide
insufficient statistics. Examples of these potentials are included in the online
supplementary information. The full energetics of the system is then described as
\begin{equation} \label{eqn:two_step}
F_{\mathrm{DFT}}=F_{\mathrm{\mathrm{an}}}+\Delta
F_{\mathrm{an} \to \mathrm{cl}}+\Delta F_{\mathrm{cl}\to \mathrm{DFT}}
\end{equation}
where $\Delta F_{\mathrm{cl}\to \mathrm{DFT}}$ requires a small number of DFT-md
simulations, and $\Delta F_{\mathrm{an} \to \mathrm{cl}}$ numerous, but inexpensive
classical Monte Carlo (CMC) simulations. The method depends on a smooth integration of
$\Delta F_{\mathrm{cl}\to \mathrm{DFT}}$ and avoiding any first order phase transitions
with $\lambda$.  We use five $\lambda$ points, for all $\Delta F_{\mathrm{cl}\to
\mathrm{DFT}}$ integrations. For solid MgO we use a combination of classical pair and
one-body harmonic oscillator potentials for $U_{\rm cl}$ \citep{Wilson2012a,Wahl2013}.
Liquids we use only pair potentials. For solids the analytical reference system is an
Einstein solid with atoms in harmonic potentials centered on a perfect lattice, while a
gas of non-interacting particles is used for liquids. We found integrating over 5 lambda
points to be sufficiently accurate, with an increase to 9 lambda points changing our
results by $<0.003$ eV per formula unit. In the online supplementary information we
include two additional tests of the thermodynamic integration method, demonstrating that
our results are not sensitive use of different classical potentials or the integration
path with respect to the interaction between different species in the multicomponent
systems.

All DFT simulations presented here were performed using the Vienna {\it ab initio}
simulation package (VASP) \citep{Kresse1996}. VASP uses projector augmented wave
pseudopotentials \citep{Blochl1994} and the exchange-correlation functional of Perdew,
Burke and Ernzerhof \citep{Perdew1996}. Although the DFT formalism is based on a
zero-temperature theory, DFT-MD simulations at high temperatures have been shown to agree
with theory developed for warm dense matter \citep{Driver2012}. We use an iron
pseudopotential with valence states described by a
$[\mathrm{Mg}]\mathrm{3pd}^6\mathrm{4s}^2$ electron configuration.  For consistency, all
simulations use Balderesci point sampling, a 600 eV cutoff energy for the plane wave
expansion and temperature dependent Fermi-smearing to determine partial orbital
occupations. A time step of between 0.5 and 1.0 fs is used depending on the temperature,
with the smaller time step used for all simulations with temperatures above 6000 K. We
confirmed that the resulting molecular dynamics results are well-converged with respect
to the energy cutoff and time step.  All presented results involve molecular
dynamics simulation 
lengths of at least 2 ps simulated time at each, with longer simulation times having an
insignificant effect on the results of the thermodynamic integration. The largest source
of uncertainty was the finite size effect, which we discuss in detail in
Section \ref{sec:results}.

The workflow for the calculation is as follows: 1) Determine the density of the system at
the target pressure and temperature by iteratively correcting the volume until the
calculated DFT pressure matches the target pressure with a 1\% tolerance. 2) Perform DFT
simulations to determine the internal energy, and fit classical potentials for the
thermodynamic integration. 3) Perform the thermodynamic integration with the classical
potentials with DFT simulations, 4) and CMC simulations to complete the Helmholtz free
energy (Eqn. \ref{eqn:two_step}), and by extension the entropy and Gibbs free energy. 5)
Calculate the differences in these values at given $P$ and $T$, using Eqn.
\ref{eqn:gibbs} to determine whether the system mixes or not.

\section{Results} \label{sec:results}

\begin{figure}[h!]  
  \centering
    \includegraphics[width=20pc]{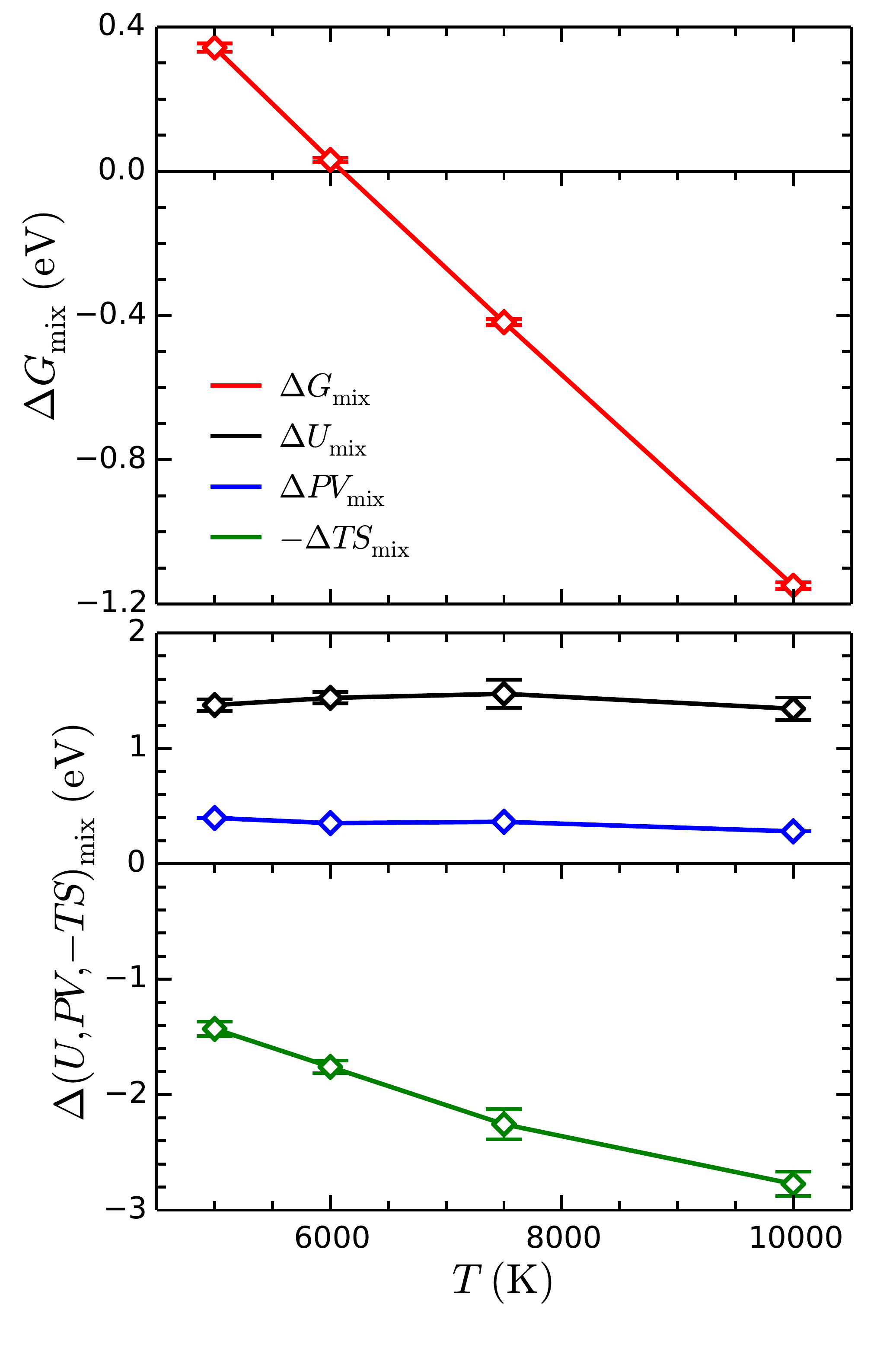}
\caption{Gibbs free energy change per formula unit, $G_{\rm{mix}}$ of the reaction
${ \rm MgO_{liq} + Fe_{liq} } \rightarrow {\rm FeMgO_{liq}}$ at $P=50$ GPa (red).  Independently
calculated components of $\Delta G_{\rm{mix}}$: $\Delta U$ (black), $\Delta PV$ (blue),
and $-\Delta TS$ (green). Positive values favor separation into end member phases,
while negative values favor a single mixed phase. $\Delta PV$ values presented here use
the target pressure. Error bars represent the integrated error from the 1 $\sigma$
statistical uncertainty of the molecular dynamics simulations.}
\label{fig:components}
\end{figure}

The online supplementary material includes a table with the results of the calculations
for each composition and $P$-$T$ condition. It includes the density along with the
calculated pressure, internal energy, entropy and Gibbs free energy. The stable phase at
each condition is determined using Eqn. \ref{eqn:gibbs}.  The point at which the trend in
$\Delta G_{\rm mix}$ at constant $P$ changes sign is the inferred solvus temperature at
the 1:1 stoichiometric composition.  Fig.  \ref{fig:components} shows an example of this
trend in $\Delta G$ for the 1:1 mixture at $P=50$ GPa, along with its components $\Delta
U$, $\Delta PV$ and $-\Delta TS$. Using the convention from Eqn. \ref{eqn:gibbs},
positive values favor the separation of the  material into the endmember phases, while
negative values favor the single homogeneous mixed phase. The contributions of the
internal energy and volumetric terms are positive, while the entropy provides the
negative contribution that promotes mixing at sufficiently high temperature. 

\begin{figure}[h!]  
  \centering
    \includegraphics[width=20pc]{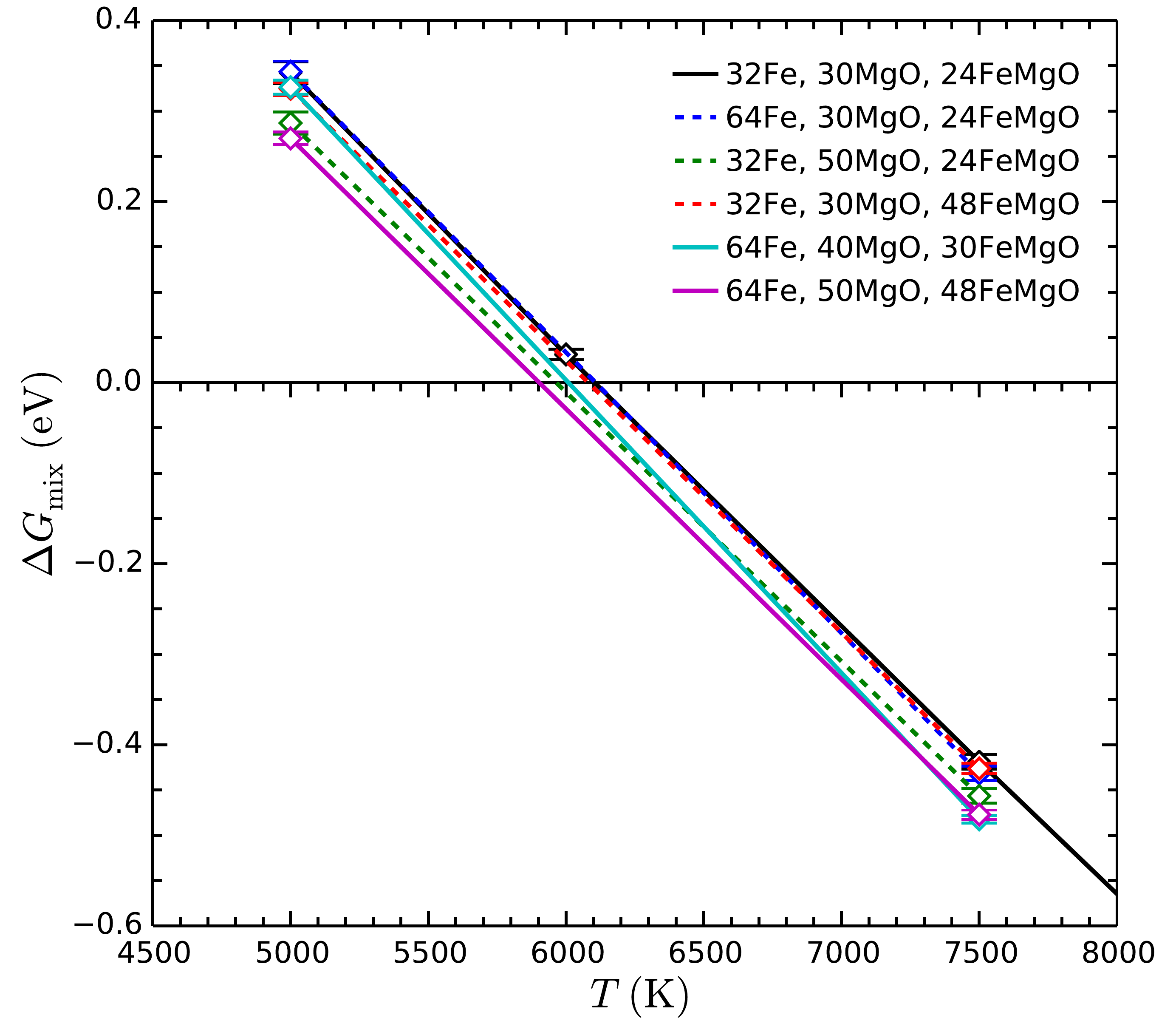}
\caption{Quantifying the finite size effect on $\Delta G_{\rm{mix}}$ for simulations of
the reaction ${ \rm MgO_{liq} + Fe_{liq} } \rightarrow {\rm FeMgO_{liq}}$ at $P=50$ GPa
with different cell sizes. In black are the results for the systems  ${\rm Fe_{32}}$,
${\rm Mg_{32}O_{32}}$ and ${\rm Fe_{24}Mg_{24}O_{24}}$ used at the other P-T conditions.
The other lines show the shift in $\Delta G_{\rm{mix}}$ obtained when the calculation is
repeated for with a larger cell for one (dashed lines) or all (solid lines) of the systems.}
\label{fig:finite_size_effect}
\end{figure}

Using additional calculations at $P=50$  GPa, we estimated uncertainties in our
calculated Gibbs free energies. The most significant contribution to the uncertainty
comes from the finite size of the simulation cells. We estimate the magnitude of this
uncertainty by comparing results from larger simulated cells to those of the original
system. Fig. \ref{fig:finite_size_effect} compares the values of $\Delta G_{\rm mix}$ for
Fe and FeMgO cells with up to twice the number of atoms, and MgO with up to 100 atoms per
cell. From this, we estimate a maximum shift of $<0.1$ eV per formula unit. This
corresponds to an uncertainty in temperature of $\sim$200 K, roughly an order of
magnitude larger than the statistical precision of the calculation. The combined effect
of increasing cell size for all systems leads consistently to lower values of $\Delta
G_{\rm mix}$ at both temperatures, and thus, lower predicted solvus closure temperatures.
For the subsequent analysis, we consider an estimated uncertainty defined as the largest
(positive or negative) shift in the Gibbs free energy for each phase. It should be noted
that these estimated error bars can not be strictly viewed as statistical uncertainties,
since they are unidirectional and based on a small number of independent calculations.
They suggest the likely magnitude by which similar shifts in the calculated Gibb's free
energies will effect the calculated transition temperature at different $P-T$ conditions. 

\begin{figure}[h!]  
  \centering
    \includegraphics[width=16pc]{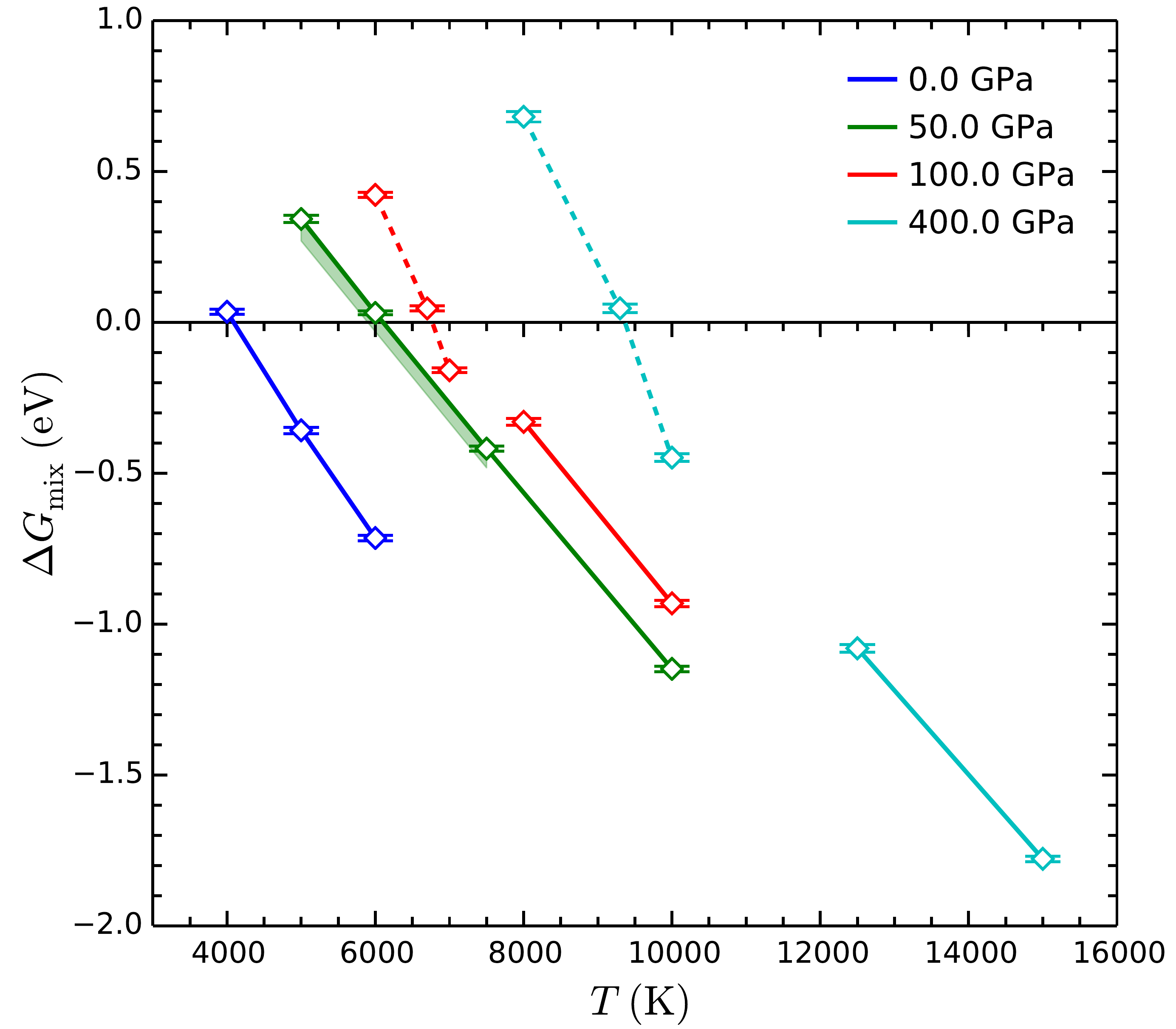}
\caption{Gibbs free energy of mixing for MgO and liquid Fe. Solid lines show conditions
  where MgO was simulated as a liquid, and dashed lines where MgO is in its (B1) solid
  phase.  The filled green region shows an estimate of the uncertainty from finite size
  effects, taken as the maximum shifts in $\Delta G_{\rm{mix}}$ observed our tests of larger
  cells (Fig. \ref{fig:finite_size_effect}).
}
\label{fig:deltaG}
\end{figure}

The effect of pressure and temperature on $\Delta G_{\rm mix}$ is shown in Fig.
\ref{fig:deltaG}. As pressure increases, the slope of $\Delta G_{\rm mix}$ with $T$
remains nearly constant for all simulations with liquid MgO, but the values are shifted to
higher temperatures. This means that the solvus closure temperature has a positive slope
with pressure over the entire range of conditions considered. There is a noticeable
change in the slope of this quantity  when MgO melts.  This corresponds with a weaker
dependence on pressure at high pressures, where the solvus temperature is below the
melting temperature of MgO. 

When determining the energetics of the mixed FeMgO, phase it is important to verify that
the simulation remain in a single mixed phase. At temperatures sufficiently close to the
solvus closure temperature the system should behave as a super-cooled homogeneous
mixture, while at sufficiently low temperatures the simulations could, in principle,
spontaneously  separate into two phases. This would bias the results as interfacial
energies between the separating phases would be included in the calculated Gibbs free
energy.  We were unable to detect phase separation by visual inspection of various of
snapshots as has been seen for hydrogen-helium mixtures in \citep{Soubiran2012}. The pair
correlation function, $g({\bf r})$, can be used as a proxy for separation of phases
\citep{Soubiran2012}.

\begin{figure}[h!]  
  \centering
    \includegraphics[width=20pc]{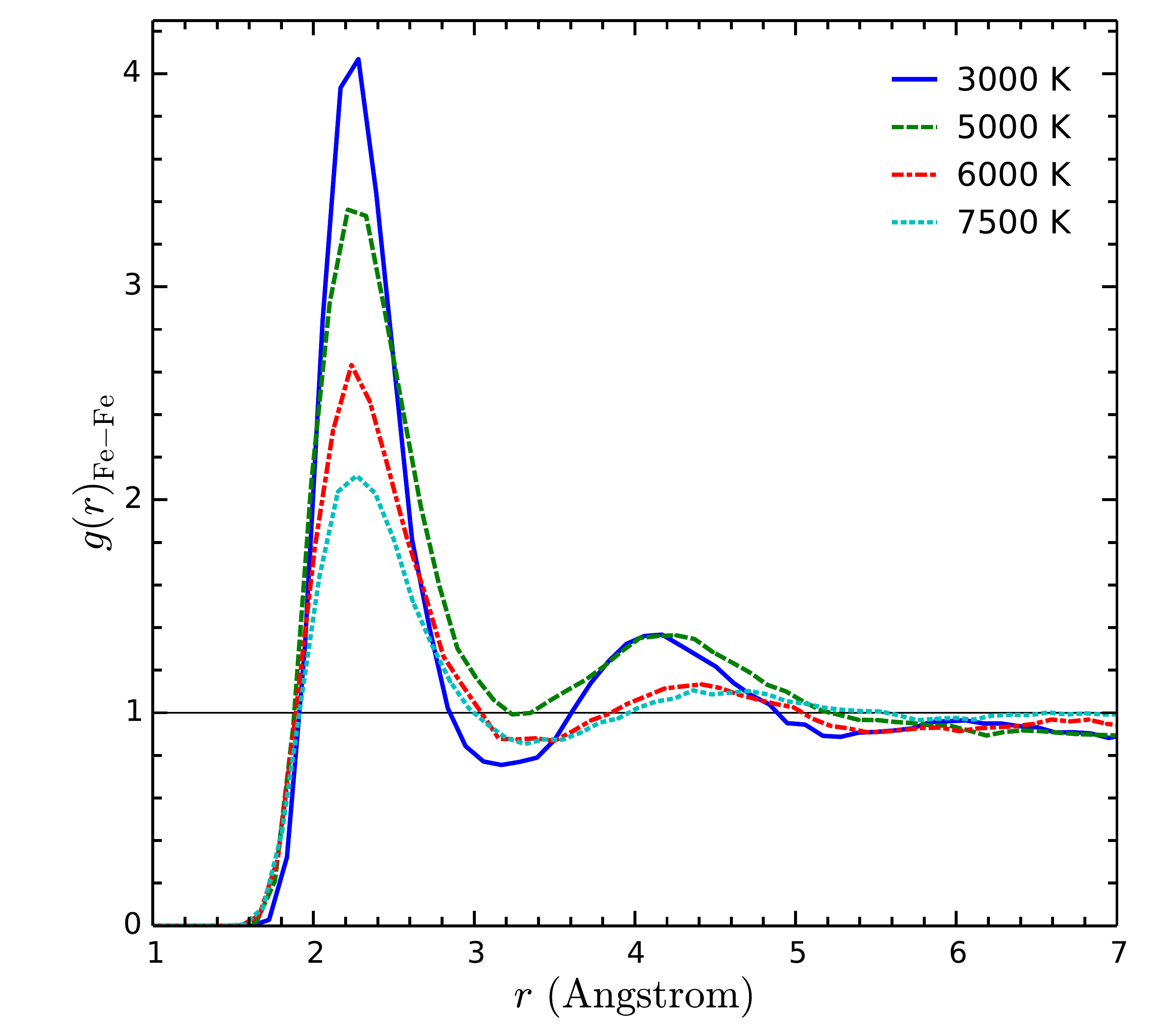}
\caption{Fe-Fe pair correlation functions for mixed Fe + MgO phase. Compares the spatial
  distribution of atoms in simulations at 50 GPa with different temperatures. The 3000 K
  and 5000 K asymptote to values notably less than one, while temperatures near or above
  the solvus closure temperature do not show such a deviation at large $r$.}
\label{fig:rdf}
\end{figure}

Fig. \ref{fig:rdf} shows the Fe-Fe $g(r)$ for simulations of the mixed FeMgO phase at 50
GPa. For temperatures significantly below solvus closure temperature, 3000 and 5000 K,
there are slight negative deviations of the $g(r)_{\rm Fe-Fe}$ at large $r$ from their
expected asymptote to unity. This is consistent with clustering into MgO and Fe-rich
regions, and may indicate spontaneous phase separation  at temperatures well below the
inferred solvus closure temperature. These deviations are minimal or not observed in
simulations near or above the inferred solvus closure temperature. We performed
additional simulations at temperatures close to the solvus closure temperature to verify
that the Gibbs free energy changes linearly as a function of temperature, which is the
expected behaviour without spontaneous phase separation. In spite of deviations in
$g(r)$, we note that including the low temperature simulations in our calculation of the
solvus closure temperature does not significantly change the result at any pressure.

\begin{figure}[h!]  
  \centering
    \includegraphics[width=20pc]{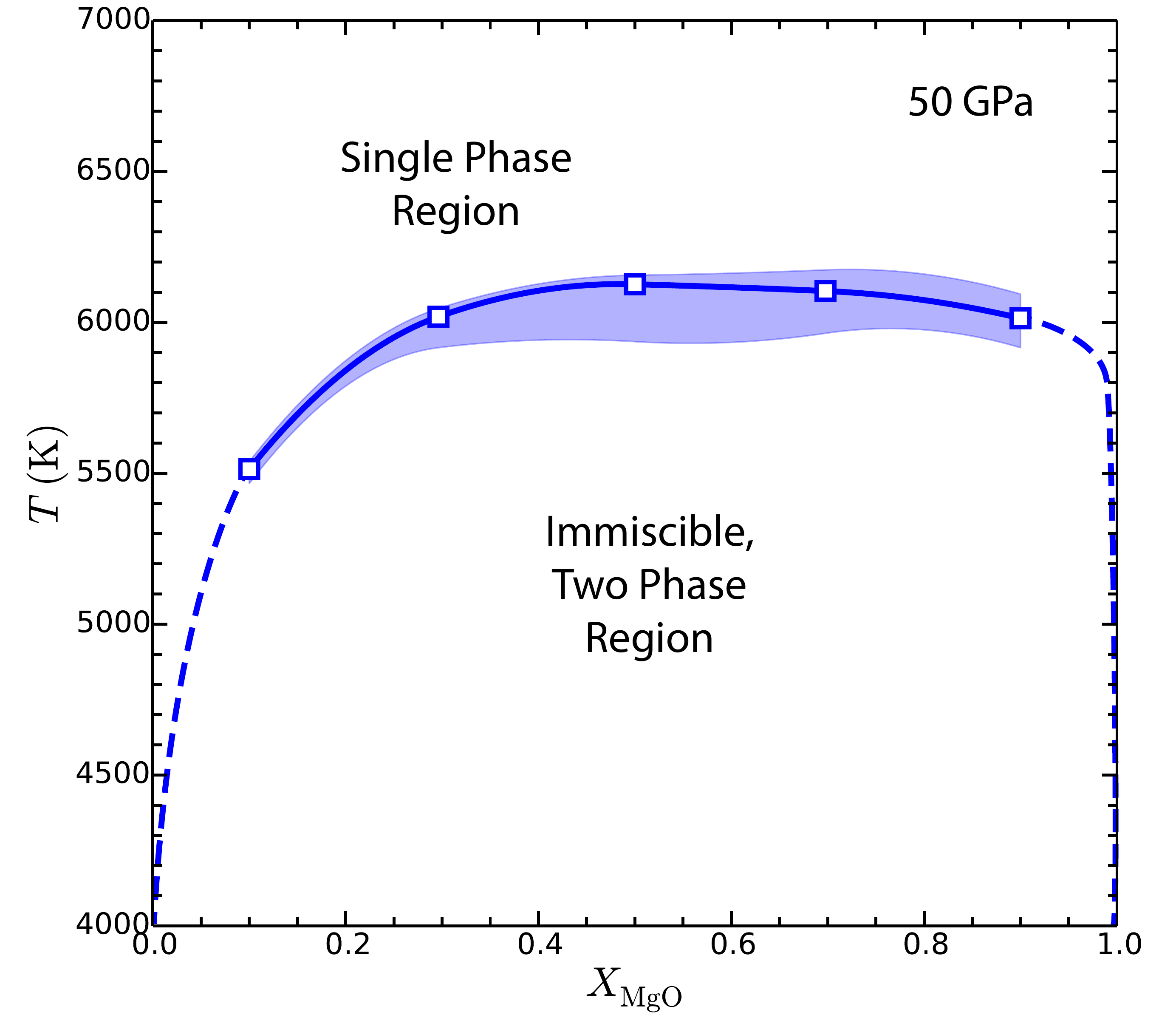}
\caption{Solvus phase diagram of the Fe-MgO system at $P=50$ GPa. The shape is consistent
  with the composition $X_{\rm MgO}=0.5$ being representative for estimating the solvus
  closure temperature at other pressures. The filled blue region shows an estimate of the
  uncertainty in transition temperature arising from the uncertainties in $G$ in Fig.
  \ref{fig:convex_hull}.}
\label{fig:solvus}
\end{figure}

\begin{figure}[h!]  
  \centering
    \includegraphics[width=16pc]{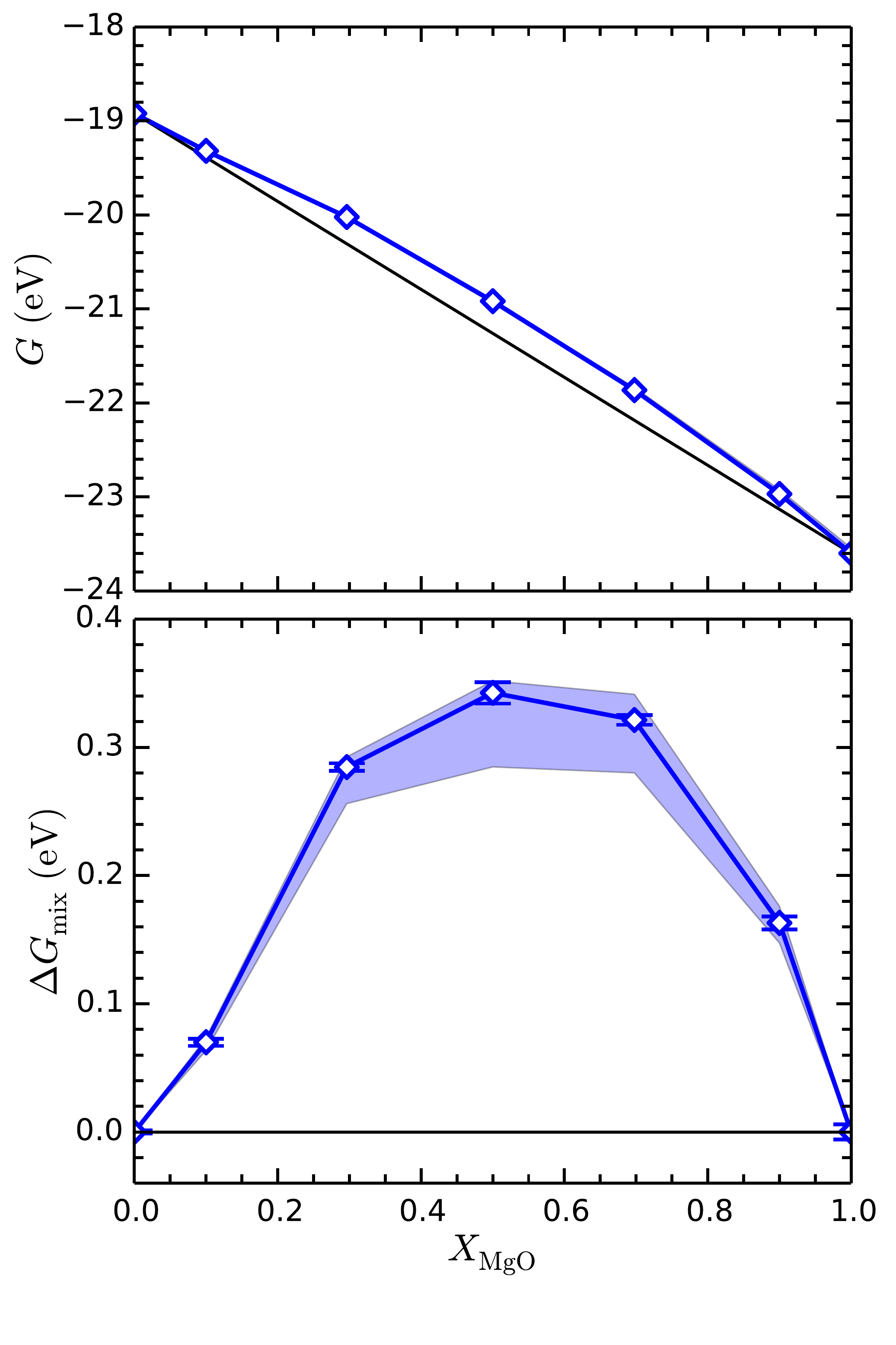}
\caption{Convex hull of $\Delta G_{\rm{mix}}$ versus formula unit fraction, $X_{\rm
MgO}$, for the Fe-MgO system at $P=50$ GPa and $T=5000$ K (top). Difference between
$\Delta G_{\rm{mix}}$ and a mixing line between the endmembers (bottom). The filled blue
region shows an estimate of the uncertainty from finite size effects, taken as the
maximum shifts in $\Delta G_{\rm{mix}}$ from our tests of larger Fe, MgO and FeMgO
simulations. The estimated error is weighted as a function of composition since the
finite size effects will cancel with that of the end-member as the compositions become
more similar.
}
\label{fig:convex_hull}
\end{figure}

We also studied the effect of composition on the solvus temperature.  Calculations were
performed on four additional intermediate compositions between the Fe and MgO endembers.
Fig. \ref{fig:convex_hull} shows a convex-hull in $G_{\rm Fe_{1-x}MgO_{x}}$ and $\Delta
G_{\rm mix}$ at 50 GPa and 5000 K. This corresponds to a temperature below the calculated
solvus temperature. The Gibbs free energies of all intermediate components are above the
mixing line between the end members, and form a smooth function with composition. This is
consistent with a binary system with a miscibility gap. Using linear interpolation
between this convex hull and one at 7500 K, we estimate the shape of the miscibility gap
at 50 GPa, as shown in Fig. \ref{fig:solvus}. The miscibility gap is notably asymmetric,
with temperatures decreasing faster towards the Fe-rich endmember than the MgO-rich end.
A similar, more pronounced asymmetry has been experimentally determined for the Fe-FeO
system at lower pressures \citep{Ohtani1984a,Kato1989}. In spite of this, the shape of the
solvus at intermediate compositions, $\sim$0.3-0.9 molar fraction MgO, is relatively
flat. As a result, the temperatures predicted for a 1:1 mixture provide a good estimate
for the solvus closure temperature. We note, however, that the shape of miscibility gap
may be sensitive to uncertainties from the finite size effect. Considering the estimated
errors from the finite size effect test, we can only constrain to be within that
$\sim$0.3-0.9 $X_{\rm MgO}$ range. Regardless of this composition our uncertainty in the
temperature of solvus closure remains $\sim200$ K.

Fig. \ref{fig:summary} summarizes the results, showing all the conditions at which
simulations were performed.  We find the solvus closure at ambient pressure to be
$4089^{+25}_{-235}$ K.  While there is little experimental work on this exact system, our
results are superficially consistent with extrapolations of the phase diagram for the
Fe-FeO system from low temperatures \citep{Ohtani1984a,McCammon1983}, and with the
`accidental' discovery of the Fe-silicate solvus by \cite{Walker1993}. We find that the
solvus temperature increases with pressure to $6010^{+28}_{-204}$ K at 50 GPa, but its
slope decreases significantly at higher pressures, with a temperatures of
$6767^{+14}_{-135}$ K at 100 GPa and $9365^{+14}_{-130}$ K at 400 GPa.  This transition
also corresponds roughly to the pressure where the trend crosses the MgO melting curve
\citep{Alfe2005, Belonoshko2010,Boates2013}.  Indeed, the simulations used to infer the
closure temperature at these pressures used the solid (B1) structure of MgO.
Unfortunately, it is difficult to check whether the change in slope is a direct result of
this phase transition, as liquid MgO simulations rapidly freeze at temperatures far below
the melting curve.  Conversely, the liquidus of a deep magma ocean might be below the
solvus at these temperatures due to the effect of an $\rm{SiO_2}$ or FeO component in the
silicate/oxide endmember \citep{deKoker2013,Zerr1998}. However, extrapolation of $\Delta
G_{\rm mix}$ from simulations with liquid MgO at higher temperatures (Fig.
\ref{fig:deltaG}) suggest that the change in slope occurs occurs in liquids as well. We
estimate shifts in the inferred solvus closure temperature from finite size effects on
the order of 200 K. The  actual solvus closure temperature may also be shifted by up to a
couple hundred Kelvin, if we also consider the uncertainty in the solvus shape (Fig.
\ref{fig:convex_hull}), since our results refer specifically the solvus temperature for a
1:1 stoichiometric mixture.  The observed change in slope of the solvus temperature and
the relation to the pure MgO melting curve are, however, robust against uncertainties of
this magnitude.

\section{Discussion} \label{sec:discussion}

\begin{figure}[h!]  
  \centering
    \includegraphics[width=20pc]{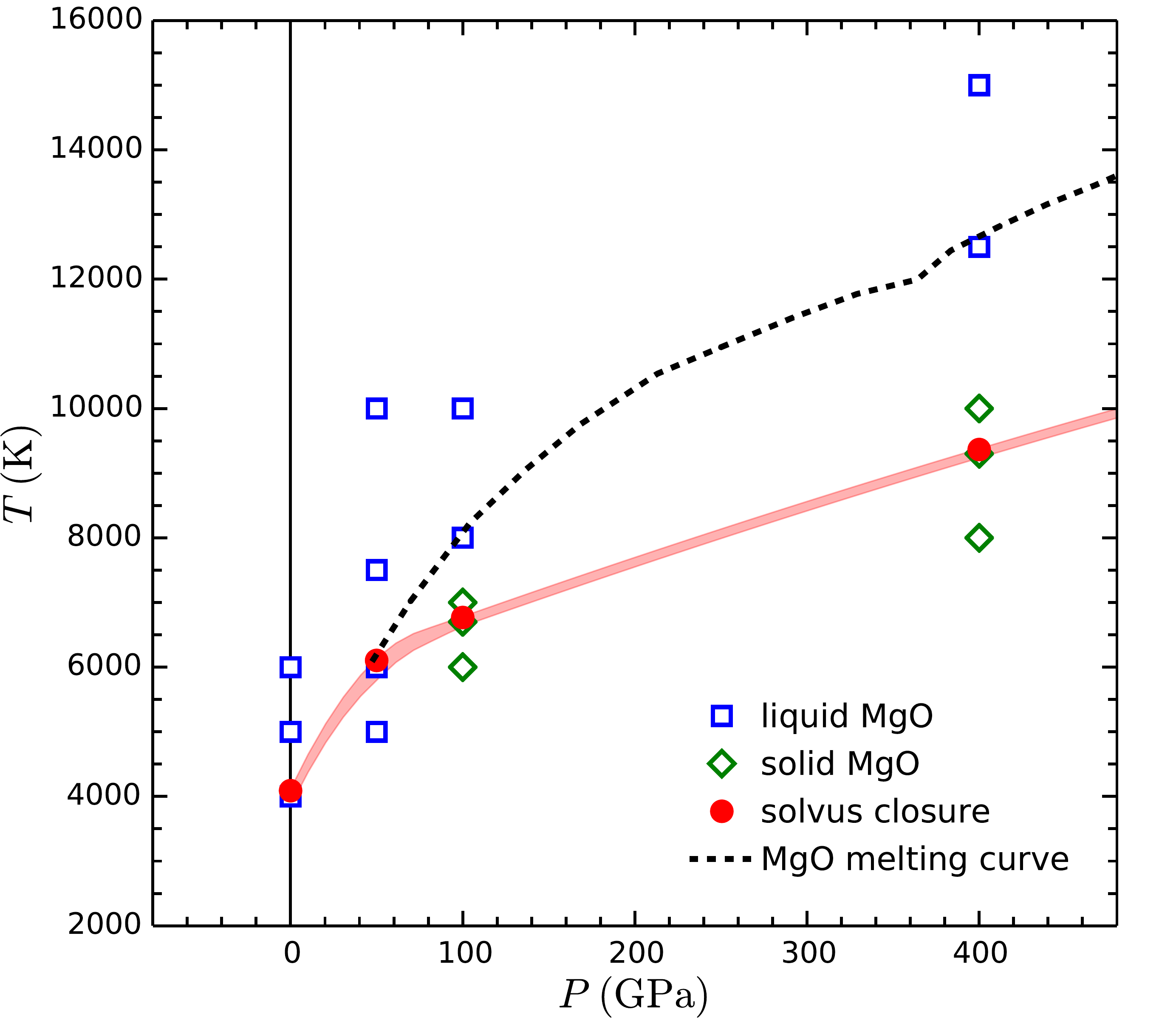}
\caption{Pressure dependence of the solvus closure. The $P$-$T$ condition of all
  thermodynamic integration calculations are included. Blue markers denote conditions
  where MgO was treated as a liquid. Green markers denote conditions where MgO was
  treated as a solid (B1).  Red circles show the solvus closure temperature inferred from
  simulations at the same pressure. The estimated uncertainty in the solvus closure
  temperature from finite size effects is shown by the filled red region.  The dashed,
  black line shows the MgO melting temperature from molecular dynamics from DFT-md with
  PBE exchange correlation function\citep{Boates2013}, which is consistent with other
  first-principles calculations \citep{Alfe2005,Belonoshko2010}}
\label{fig:summary}
\end{figure}

In the extreme case where a significant fraction of the planet is in a mixed iron-rock
phase, the early evolution will be quite different than prevailing theories.
Differentiation of material accreted onto the planet is delayed until the planet cools to
below the solvus closure temperature, allowing iron to exsolve and sink to the core.
This study provides an estimate of the temperatures required to mix the Mg-rich rocky
mantle with the core of a terrestrial planet. At the surface, the complete mixing of Fe
and MgO is achieved at $\sim$4000 K (Fig. \ref{fig:summary}), which is well above the
melting point of silicates. At core-mantle boundary pressures, the critical temperature
would be $\sim$7000 K.  This is below higher estimates for the melting temperature of
pure MgO \citep{Alfe2005,Belonoshko2010,Boates2013} and $\rm{MgSiO_3}$ perovskite
\citep{Zerr1993}. There are significant disparities among calculations and experiments
on the melting \citep{Belonoshko1994,Belonoshko1997,Alfe2005,Zerr1998}
temperatures in the lower mantle, disagreeing even on which phases represent the solidus
and liquidus. The melting behavior in our MgO simulations are consistent with the
high-temperature melting curve of recent first-principles simulations
\citep{Alfe2005,Belonoshko2010,Boates2013}.  Regardless, the solvus remains well above
the solidus for more realistic compositions of the silicate mantle
\citep{deKoker2013,Zerr1998,Holland1997}.

\subsection{Evolution of a fully mixed planet}

For a sufficiently energetic impact, or series of impacts, a planet might be heated to
such high temperatures, that the entire planet maybe be an approximately homogeneous
mixture of the iron and rock components. Such an extreme scenario is unlikely for an
Earth-sized planet, and likely violates geochemical observations that preclude complete
mixing of the Earth's primitive mantle \citep{Mukhopadhyay2012}. Nonetheless, considering the
evolution of a planet from a fully-mixed state is useful for demonstrating the effects
our phase diagram on the mixing behavior in a planet. A fully mixed state is also not so
far-fetched for super-Earths since heating from release of gravitational energy scales as
roughly $M^{2/3}$. 

The depth at which the phases separate from the fully mixed state is determined by the
pressure dependence of solvus closure. Following such a large impact, the planet will
quickly evolve to a magma ocean state, and a higher-temperature adiabat will be rapidly
re-established. Fig. \ref{fig:isentropes} compares the solvus closure temperature to the
calculated isentropes of the mixed FeMgO phase. For a homogeneous, vigorously convecting
liquid layer of the planet, these approximate adiabatic temperature profiles of the
interior of the planet at different points in its evolution.  The comparison is
qualitatively the same if the isentropes for either endmember is used instead of the
mixed phase. At pressures above 50 GPa, the isentropes have a notably steeper slope than
the solvus closure temperature. At lower pressures, $<$50 GPa the slopes are identical within the
estimated uncertainty. As a result, separation begins in the exterior of the planet and
proceeds inwards as the planet cools. Since iron separating in the outer portion of the
planet is denser than the rocky phase, it would sink until it reached a depth where it
dissolves into the mixed phase again. This may promote compositional stratification, and
possibly multi-layer convection between an upper iron-poor and deeper iron-rich layer.
The extent to which this process can stratify the planet depends on the competition
between growth of liquid Fe droplets and their entrainment in convective flows
\citep{Solomatov2007}.

Based on the Fe-MgO solvus closure temperature presented here, transition of a planet
from a fully mixed state to separated rocky and metallic phases would occur while the entire
planet is at least partially molten. Consequently, a fully mixed state in an Earth-mass
planet would be short lived, since cooling timescales for a deep magma ocean are short in
comparison to the timescale of accretion \citep{Abe1997,Elkins-Tanton2012}. This also
means little record of such an event is likely to survive to the present day Earth. Indeed,
there is little unambiguous evidence for a magma ocean, despite it being a seemingly
unavoidable consequence of the moon-forming impact hypothesis. The high surface
temperatures of some rocky exoplanets \citep{Pepe2013} might allow for prolonged cooling
times from a such a mixed state.

At the relatively low pressures of growing planetesimals
\citep{Kleine2002,Sramek2012,Weiss2013}, these results predict that core formation begins
at temperatures well below the solvus. As a result, complete mixing of a planet must
overcome the gravitationally stable differentiated structure. This will impede upward
mixing of a dense core even at temperatures above solvus closure, leading to an inefficient
double-diffusive convection state like that proposed for the giant planets
\citep{Chabrier2007}. Material accreted while the planet was above the solvus would,
however, remain in a fully mixed outer layer, and evolve according to the picture
presented in Fig. \ref{fig:isentropes}. Substantial mechanical mixing during giant impact
events \citep{Canup2004,Canup2012,Cuk2012} would also enhance mixing prior to the setup of
a double-diffusive state.

\subsection{Consequences of localized heating}

Despite the implausibility of a fully-mixed earth, related processes may become important
as material is added by impacts with smaller differentiated bodies at temperatures near
or above solvus closure. Since peak shock temperatures are related to the velocity of the
impact rather than the size of the impactor, smaller-scale events can create localized
regions where the temperature exceeds the solvus closure temperature.  Assuming iron from
the shocked region can be rapidly delivered to the core without significant cooling
\citep{Monteux2009}, material equilibrated near or above the solvus can be delivered to
the core, through a mantle of lower average temperature. In the case that heat transfer
from the sinking iron diapir is negligible, the comparison between the solvus closure
temperature and adiabats is valid for the fraction of the planet in contact with the
sinking iron. In other words, the temperature in the sinking iron will follow a nearly
adiabatic path, with Fe and MgO becoming more soluble as the pressure rises. This means
that a fraction of the iron delivered to the core could have equilibrated with the rocky
mantle at much higher temperatures than on average. The differentiation of a fraction of
the planet in the presence of a mixed phase would likely effect partitioning of both major
and minor elements between the core and mantle. \cite{Walker1993}
suggested that deviations in siderophile element partitioning behavior occur near the
solvus closure temperature for iron-silicate mixtures. However, this interpretation has
been questioned in light of the confounding effect of drastic changes in oxygen partitioning
with pressure \citep{Frost2010}. Better characterization of element-partitioning at such
high temperatures could constrain what fraction of the mantle could have been
equilibrated in this fashion.

\begin{figure}[h!]  
  \centering
    \includegraphics[width=20pc]{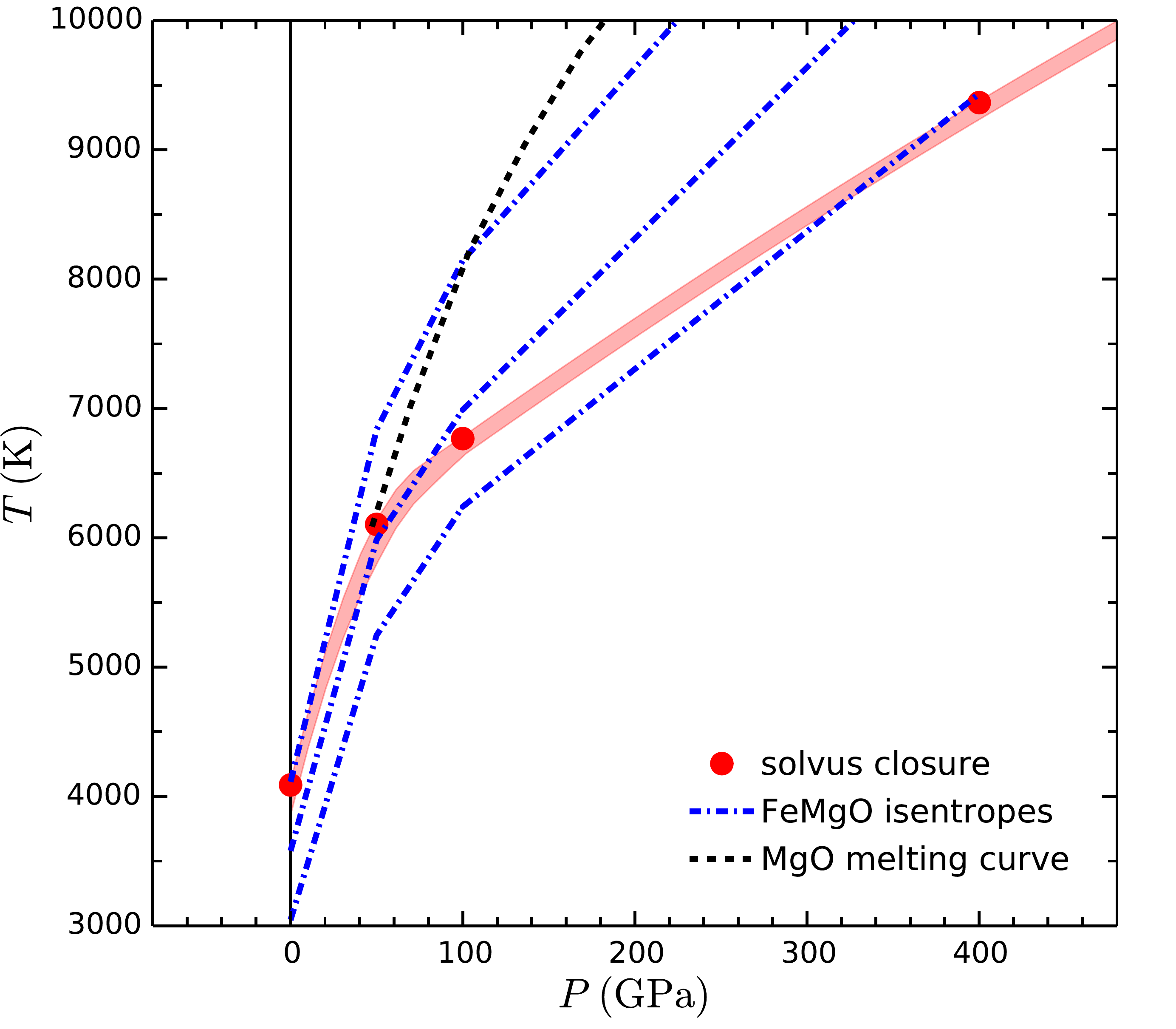}  
\caption{Calculated isentropes for the mixed FeMgO liquid phase compared to the solvus
  closure temperature. These results favor the mixed phase to remain stable at depth. The
  dashed, black line shows the MgO melting temperature from \citep{Boates2013}.The filled
  regions represent the propagation of estimated errors from finite size effects.}
\label{fig:isentropes}
\end{figure}

\begin{figure}[h!]  
  \centering
    \includegraphics[width=20pc]{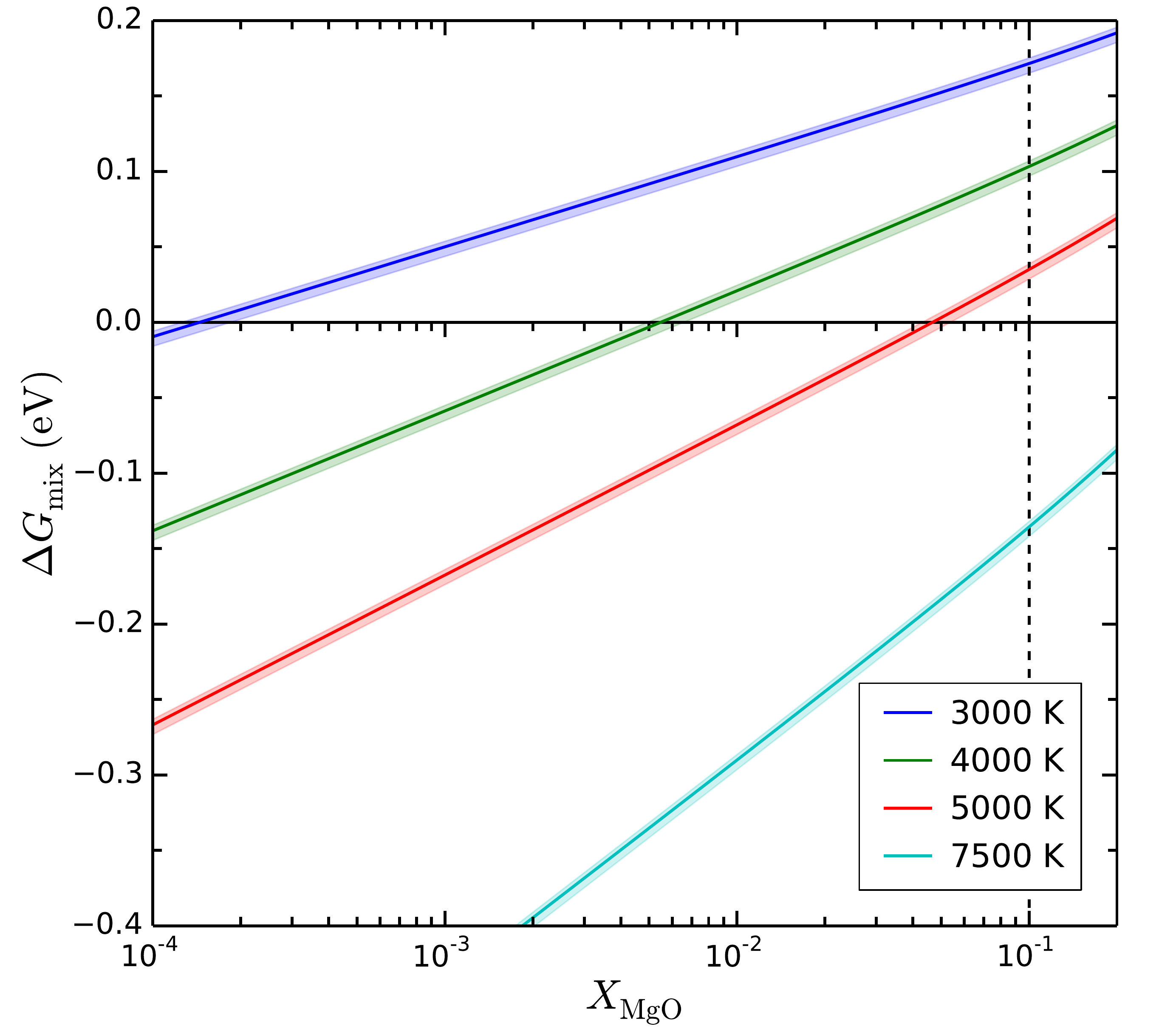}  
\caption{Extrapolated saturation limits of MgO in Fe at 50 GPa at various temperatures.
  Extrapolation is under the assumption that the solution behaves in the
  low-concentration limit. The dashed vertical line is the most Fe-rich composition from
  Fig. \ref{fig:convex_hull}, from which the extrapolation is made. The filled regions
  represent the propagation of estimated errors from finite size effects.}
\label{fig:concentrations}
\end{figure}

One important consequence of high-temperature equilibration is the delivery of excess,
nominally insoluble, light components to the core. This will occur if iron is
equilibrated with rocky materials at near-solvus temperatures, and rapidly delivered to
the core before it can cool and re-equilibrate with the mantle at lower temperatures.
This would be followed by exsolution of a Mg-rich material at the top of the cooling
core. This process has been suggested as a possible solution to the problem of the
Earth's core having  insufficient energy to generate a magnetic field before nucleation
of the inner core \citep{Stevenson2012}. If the interpretation of Fig.
\ref{fig:isentropes} can be extended to more iron rich compositions, then exsolution will
occur at the top of the core, depositing sediments of Mg-rich material at the core-mantle
boundary \citep{buffett2000}. As a result, the effect of this sedimentation on core
convection is analogous to the exclusion of light elements from the growing inner core.
Fig. \ref{fig:concentrations} shows an extrapolation of our results to predict the
saturation of MgO in Fe at 50 GPa as a function of temperature.  This is done using a
function for $G$ in terms of the cell volumes derived in the low-concentration limit
\citep{Wilson2012a,Wahl2013}, and details are presented in the supplemental material. From
this we predict a $>$1\% MgO saturation limit down to 4200 K, with concentrations steeply
decreasing to be below detection limits in high-pressure experiments by $\sim$3000 K
\citep{Knittle1991,Ozawa2008a}. In principle, high-temperature equilibration could also
explain a bulk-mantle iron concentration in disequilibrium with the present-day core
\citep{Ozawa2008a,Tsuno2013}. However, the shape of the MgO-rich side of the calculated
exsolution gap (Fig. \ref{fig:solvus}) contradicts this, since the Fe content of the MgO
endmember shows a significant deviation for only a small range of temperatures.

\section{Conclusions}

The solvus closure temperatures for material with the bulk terrestrial planet composition
marks the transition to a regime where where miscibility is a dominant effect in the
evolution of the planet. These results present an estimate of those temperatures based on
the simplified Fe-MgO system. Where possible, our simulated system was chosen to provide
an upper limit for these temperatures, so we expect miscibility for realistic terrestrial
compositions at possibly lower temperatures than those found here. The solvus closure
temperature found here for the Fe-MgO system is at temperatures low enough, that it was
likely overcome for some fraction of the planet during accretion. Energetic impact events
are now thought to have been commonplace during the formation of the terrestrial planets,
and the role of miscibility between the most abundant rocky and metallic materials should
be considered to adequately assess their early evolution.

\section*{Acknowledgements}

This work was supported by NASA and NSF. Computational resources were provided in part by
the NASA Advanced Supercomputing Division.  We thank Dave Stevenson, Raymond Jeanloz, 
Sarah Stewart and James Brenan for helpful discussions.

\section*{References}

\bibliography{femgo}




\end{document}


\begin{frontmatter}

  \title{High-temperature miscibility of iron and rock during terrestrial planet
  formation}



\author[mymainaddress]{Sean M.Wahl\corref{mycorrespondingauthor} }
\cortext[mymainaddress,mycorrespondingauthor]{Corresponding author}
\ead{swahl@berkeley.edu}

\author[mymainaddress,mysecondaryaddress]{Burkhard Militzer}

\address[mymainaddress]{Department of Earth and Planetary Science, University of California
Berkeley, United States}
\address[mysecondaryaddress]{Department of Astronomy, University of California
Berkeley, United States }

\end{frontmatter}

\section{Supplementary material}

\subsection{Simulation results} \label{sec:simulation}
Table~\ref{tab:results} shows the results of thermodynamic integration
calculations performed for each composition and $P$-$T$ condition. It
includes the density along with the calculated pressure, internal energy
$U$, entropy $S$, and Gibbs free energy, $G$. Calculations are specified by
pressure, $P$, temperature, $T$, and the atomic composition and phase.

To determine the density $\rho$ for a given $P$ and $T$ of interest, we
fitted equations of state to results from DFT-MD simulations. $P$ and $U$
are time-averaged results from DFT-MD simulations with 1$\sigma$
statistical error quoted. $S$ and $G$ are calculated from the two step
thermodynamic integration technique. For quantities calculated using the
thermodynamic integration, the quoted errors were derived by propagating
the errors from each integration point.

\begin{deluxetable}{ccllllll}
\tabletypesize{\scriptsize}
\tablecolumns{8}
\tablewidth{0pc}
\tablecaption{Thermodynamic functions derived from DFT-MD simulations. 
\label{tab:results}}

 \tablehead{ \colhead{$P$} & \colhead{$T$} & \colhead{system} &  \colhead{$\rho$} & 
 \colhead{$P$} & \colhead{$U$} & \colhead{$S$} &  \colhead{$G$}  \\
 \colhead{(GPa)} & \colhead{(K)} & &  \colhead{($\mathrm{g}/\mathrm{cm}^3$)} 
   & \colhead{(GPa)} & \colhead{(eV)} & \colhead{($k_B$)} & \colhead{(eV)} }
\startdata
             0 &   4000 &  $\rm{Mg}_{24}$$\rm{Fe}_{24}$$\rm{O}_{24}$,liq &   2.861 &   $-$0.41(17) &     $-$323(1) &    1064(4) &     $-$689.3(1) \\
             . &      . &                             $\rm{Fe}_{32}$,liq &   6.885 &    0.06(21) &  $-$206.53(9) &   496.7(4) &    $-$377.74(3) \\
             . &      . &                $\rm{Mg}_{32}$$\rm{O}_{32}$,liq &   2.003 &    0.60(14) &   $-$275.0(5) &     776(2) &    $-$542.40(5) \\
             0 &   5000 &  $\rm{Mg}_{24}$$\rm{Fe}_{24}$$\rm{O}_{24}$,liq &   2.556 &    0.15(16) &     $-$289(1) &    1152(3) &     $-$785.4(2) \\
             . &      . &                             $\rm{Fe}_{32}$,liq &   6.340 &   $-$0.05(34) &   $-$189.8(1) &   540.1(4) &    $-$422.52(2) \\
             . &      . &                $\rm{Mg}_{32}$$\rm{O}_{32}$,liq &   1.660 &    0.10(10) &   $-$242.8(8) &     860(2) &    $-$613.17(6) \\
             0 &   6000 &  $\rm{Mg}_{24}$$\rm{Fe}_{24}$$\rm{O}_{24}$,liq &   2.195 &   0.639(88) &   $-$251.9(8) &    1231(2) &     $-$888.3(1) \\
             . &      . &                             $\rm{Fe}_{32}$,liq &   5.740 &   $-$0.02(12) &   $-$170.8(1) &   580.2(3) &    $-$470.80(3) \\
             . &      . &                $\rm{Mg}_{32}$$\rm{O}_{32}$,liq &   1.435 &   0.707(49) &     $-$213(1) &     924(2) &    $-$690.65(9) \\
            50 &   5000 &  $\rm{Mg}_{24}$$\rm{Fe}_{24}$$\rm{O}_{24}$,liq &   5.237 &   49.28(40) &   $-$315.5(7) &     963(2) &     $-$502.0(2) \\
             . &      . &                             $\rm{Fe}_{32}$,liq &   8.805 &   51.36(34) &   $-$203.3(3) &   475.0(6) &    $-$302.73(2) \\
             . &      . &                $\rm{Mg}_{32}$$\rm{O}_{32}$,liq &   3.582 &   51.41(26) &   $-$261.4(5) &     703(1) &    $-$377.59(9) \\
             . &      . &    $\rm{Mg}_{6}$$\rm{Fe}_{54}$$\rm{O}_{6}$,liq &   7.797 &   50.61(43) &   $-$370.3(5) &     988(1) &    $-$579.56(9) \\
             . &      . &  $\rm{Mg}_{16}$$\rm{Fe}_{38}$$\rm{O}_{16}$,liq &   6.347 &   49.70(24) &   $-$338.0(8) &     995(2) &    $-$540.61(8) \\
             . &      . &  $\rm{Mg}_{30}$$\rm{Fe}_{13}$$\rm{O}_{30}$,liq &   4.461 &   49.73(22) &   $-$302.3(7) &     911(2) &    $-$470.07(8) \\
             . &      . &   $\rm{Mg}_{36}$$\rm{Fe}_{4}$$\rm{O}_{36}$,liq &   3.838 &    47.6(12) &     $-$317(3) &     855(8) &     $-$459.4(1) \\
             . &      . &  $\rm{Mg}_{30}$$\rm{Fe}_{30}$$\rm{O}_{30}$,liq &   5.237 &   47.94(21) &     $-$400(1) &    1190(3) &    $-$627.26(8) \\
             . &      . &  $\rm{Mg}_{48}$$\rm{Fe}_{48}$$\rm{O}_{48}$,liq &   5.237 &   48.49(21) &     $-$640(1) &    1908(3) &    $-$1004.9(2) \\
             . &      . &                             $\rm{Fe}_{64}$,liq &   8.805 &   50.57(18) &   $-$407.6(3) &   947.6(6) &    $-$605.53(2) \\
             . &      . &                $\rm{Mg}_{40}$$\rm{O}_{40}$,liq &   3.582 &    40.2(11) &     $-$365(3) &     788(8) &     $-$470.9(2) \\
             . &      . &                $\rm{Mg}_{50}$$\rm{O}_{50}$,liq &   3.582 &   44.43(37) &     $-$438(2) &    1023(4) &     $-$587.2(2) \\
            50 &   6000 &  $\rm{Mg}_{24}$$\rm{Fe}_{24}$$\rm{O}_{24}$,liq &   5.083 &   49.36(25) &   $-$291.5(7) &    1027(2) &     $-$587.3(1) \\
             . &      . &                             $\rm{Fe}_{32}$,liq &   8.537 &   50.95(21) &   $-$191.6(1) &   506.5(3) &    $-$345.03(2) \\
             . &      . &                $\rm{Mg}_{32}$$\rm{O}_{32}$,liq &   3.446 &   49.50(24) &   $-$243.1(5) &     754(1) &    $-$439.09(4) \\
            50 &   7500 &  $\rm{Mg}_{24}$$\rm{Fe}_{24}$$\rm{O}_{24}$,liq &   4.868 &   50.09(41) &     $-$254(2) &    1110(3) &     $-$725.7(1) \\
             . &      . &                             $\rm{Fe}_{32}$,liq &   8.164 &   50.59(52) &   $-$173.9(3) &   545.9(5) &    $-$413.29(2) \\
             . &      . &                $\rm{Mg}_{32}$$\rm{O}_{32}$,liq &   3.301 &   51.09(60) &     $-$212(1) &     823(2) &    $-$540.85(6) \\
             . &      . &    $\rm{Mg}_{6}$$\rm{Fe}_{54}$$\rm{O}_{6}$,liq &   7.221 &   49.04(17) &   $-$317.2(3) &  1119.5(5) &    $-$806.98(5) \\
             . &      . &  $\rm{Mg}_{16}$$\rm{Fe}_{38}$$\rm{O}_{16}$,liq &   5.867 &   49.54(29) &   $-$278.9(8) &    1142(1) &    $-$772.40(6) \\
             . &      . &  $\rm{Mg}_{30}$$\rm{Fe}_{13}$$\rm{O}_{30}$,liq &   4.115 &   49.90(27) &   $-$243.4(7) &    1058(1) &    $-$683.67(8) \\
             . &      . &   $\rm{Mg}_{36}$$\rm{Fe}_{4}$$\rm{O}_{36}$,liq &   3.538 &   50.10(17) &   $-$250.7(7) &    1020(1) &    $-$664.90(8) \\
             . &      . &  $\rm{Mg}_{30}$$\rm{Fe}_{30}$$\rm{O}_{30}$,liq &   4.868 &   49.80(16) &     $-$321(1) &    1382(2) &    $-$907.77(9) \\
              . &      . &  $\rm{Mg}_{48}$$\rm{Fe}_{48}$$\rm{O}_{48}$,liq &   4.868 &   50.13(15) &     $-$507(1) &    2222(2) &    $-$1451.7(2) \\
              . &      . &                             $\rm{Fe}_{64}$,liq &   8.164 &   50.62(24) &   $-$347.7(3) &  1090.7(5) &    $-$825.75(2) \\
              . &      . &                $\rm{Mg}_{40}$$\rm{O}_{40}$,liq &   3.301 &   50.75(23) &   $-$264.9(7) &    1026(1) &    $-$674.98(4) \\
              . &      . &                $\rm{Mg}_{50}$$\rm{O}_{50}$,liq &   3.301 &   51.40(27) &     $-$330(1) &    1284(2) &    $-$843.19(7) \\
\enddata
\end{deluxetable}

\begin{deluxetable}{ccllllll}
\tabletypesize{\scriptsize}
\tablecolumns{8}
\tablewidth{0pc}
\tablenum{1}
\tablecaption{(cont'd)}

\tablehead{ \colhead{$P$} & \colhead{$T$} & \colhead{system} &  \colhead{$\rho$} & 
\colhead{$P$} & \colhead{$U$} & \colhead{$S$} &  \colhead{$G$}  \\
\colhead{(GPa)} & \colhead{(K)} & &  \colhead{($\mathrm{g}/\mathrm{cm}^3$)} 
& \colhead{(GPa)} & \colhead{(eV)} & \colhead{($k_B$)} & \colhead{(eV)} }
\startdata
            50 &  10000 &  $\rm{Mg}_{24}$$\rm{Fe}_{24}$$\rm{O}_{24}$,liq &   4.546 &   51.31(32) &     $-$197(1) &    1210(2) &     $-$976.4(1) \\
             . &      . &                             $\rm{Fe}_{32}$,liq &   7.536 &   49.92(47) &   $-$142.0(2) &   601.3(3) &    $-$537.28(3) \\
             . &      . &                $\rm{Mg}_{32}$$\rm{O}_{32}$,liq &   3.054 &   50.64(51) &     $-$163(1) &     909(2) &     $-$727.8(1) \\
           100 &   6000 &  $\rm{Mg}_{24}$$\rm{Fe}_{24}$$\rm{O}_{24}$,liq &   6.144 &  100.54(60) &     $-$264(1) &     969(3) &     $-$375.5(1) \\
             . &      . &                             $\rm{Fe}_{32}$,liq &   9.771 &   97.45(69) &   $-$188.0(5) &     476(1) &    $-$244.43(4) \\
             . &      . &                $\rm{Mg}_{32}$$\rm{O}_{32}$,sol &   4.432 &  100.04(18) &   $-$261.7(3) &   598.8(7) &    $-$269.70(8) \\
           100 &   6700 &  $\rm{Mg}_{24}$$\rm{Fe}_{24}$$\rm{O}_{24}$,liq &   6.042 &   98.34(38) &   $-$254.2(7) &     998(1) &     $-$434.5(1) \\
             . &      . &                             $\rm{Fe}_{32}$,liq &   9.660 &   99.69(82) &   $-$179.6(5) &     495(1) &    $-$273.67(4) \\
             . &      . &                $\rm{Mg}_{32}$$\rm{O}_{32}$,sol &   4.381 &  100.46(35) &   $-$250.5(6) &     627(1) &    $-$307.19(8) \\
           100 &   7000 &  $\rm{Mg}_{24}$$\rm{Fe}_{24}$$\rm{O}_{24}$,liq &   6.027 &  101.60(35) &   $-$238.4(8) &    1027(1) &     $-$461.2(1) \\
             . &      . &                             $\rm{Fe}_{32}$,liq &   9.611 &  100.24(30) &   $-$176.2(2) &   502.5(3) &    $-$286.64(2) \\
             . &      . &                $\rm{Mg}_{32}$$\rm{O}_{32}$,sol &   4.360 &  100.53(42) &   $-$245.4(3) &   637.3(6) &    $-$323.25(7) \\
           100 &   8000 &  $\rm{Mg}_{24}$$\rm{Fe}_{24}$$\rm{O}_{24}$,liq &   5.932 &  103.41(68) &     $-$218(1) &    1068(2) &     $-$551.3(2) \\
             . &      . &                             $\rm{Fe}_{32}$,liq &   9.388 &   99.43(71) &   $-$165.0(5) &   527.1(8) &    $-$331.12(4) \\
             . &      . &                $\rm{Mg}_{32}$$\rm{O}_{32}$,liq &   4.034 &  100.57(67) &   $-$183.8(7) &     785(1) &    $-$393.43(7) \\
           100 &  10000 &  $\rm{Mg}_{24}$$\rm{Fe}_{24}$$\rm{O}_{24}$,liq &   5.664 &  100.86(98) &     $-$176(2) &    1147(2) &     $-$742.5(2) \\
             . &      . &                             $\rm{Fe}_{32}$,liq &   8.998 &   98.79(49) &   $-$142.5(4) &   567.2(4) &    $-$425.41(2) \\
             . &      . &                $\rm{Mg}_{32}$$\rm{O}_{32}$,liq &   3.842 &   99.23(52) &     $-$145(2) &     856(2) &    $-$534.78(8) \\
           100 &  15000 &  $\rm{Mg}_{24}$$\rm{Fe}_{24}$$\rm{O}_{24}$,liq &   5.157 &  101.17(74) &      $-$75(2) &    1281(2) &  $-$1267.13(10) \\
             . &      . &                             $\rm{Fe}_{32}$,liq &   8.145 &  102.48(80) &    $-$81.2(5) &   645.5(4) &    $-$688.15(2) \\
           400 &   8000 &  $\rm{Mg}_{24}$$\rm{Fe}_{24}$$\rm{O}_{24}$,liq &   9.117 &  398.69(56) &      $-$24(1) &     921(2) &      390.9(2) \\
             . &      . &                             $\rm{Fe}_{32}$,liq &  13.469 &   397.7(10) &    $-$93.1(8) &     453(2) &      145.0(2) \\
             . &      . &                $\rm{Mg}_{32}$$\rm{O}_{32}$,sol &   6.487 &  399.95(25) &    $-$69.8(3) &   580.4(5) &      354.3(1) \\
           400 &   9300 &  $\rm{Mg}_{24}$$\rm{Fe}_{24}$$\rm{O}_{24}$,liq &   9.010 &  400.51(49) &        5(2) &     974(2) &      285.7(2) \\
             . &      . &                             $\rm{Fe}_{32}$,liq &  13.583 &  398.04(53) &    $-$95.6(5) &   445.8(7) &      92.59(8) \\
             . &      . &                $\rm{Mg}_{32}$$\rm{O}_{32}$,sol &   6.434 &  400.24(54) &    $-$51.3(5) &   615.1(8) &      286.8(1) \\
           400 &  10000 &  $\rm{Mg}_{24}$$\rm{Fe}_{24}$$\rm{O}_{24}$,liq &   8.954 &  399.46(67) &       17(2) &     998(2) &      225.5(2) \\
             . &      . &                             $\rm{Fe}_{32}$,liq &  13.263 &   397.6(25) &      $-$75(2) &     485(2) &       65.3(1) \\
             . &      . &                $\rm{Mg}_{32}$$\rm{O}_{32}$,sol &   6.393 &  397.31(43) &    $-$41.8(3) &   632.4(5) &      249.7(1) \\
           400 &  12500 &  $\rm{Mg}_{24}$$\rm{Fe}_{24}$$\rm{O}_{24}$,liq &   8.753 &  399.86(65) &       63(1) &    1071(1) &        2.5(2) \\
             . &      . &                             $\rm{Fe}_{32}$,liq &  12.938 &   400.3(12) &    $-$46.0(9) &   530.1(9) &     $-$44.36(5) \\
             . &      . &                $\rm{Mg}_{32}$$\rm{O}_{32}$,liq &   6.100 &  396.44(75) &       61(1) &     794(1) &       82.3(1) \\
           400 &  15000 &  $\rm{Mg}_{24}$$\rm{Fe}_{24}$$\rm{O}_{24}$,liq &   8.553 &   399.7(11) &      117(2) &    1138(2) &     $-$235.9(1) \\
             . &      . &                             $\rm{Fe}_{32}$,liq &  12.688 &  403.62(99) &    $-$19.3(7) &   562.1(6) &    $-$161.93(3) \\
             . &      . &                $\rm{Mg}_{32}$$\rm{O}_{32}$,liq &   5.943 &   396.6(11) &      114(2) &     858(2) &      $-$95.8(1) \\
\enddata
\end{deluxetable}

Simulations of $\rm{Fe}_{32}$, $\rm{Mg}_{32}$$\rm{O}_{32}$ and
$\rm{Mg}_{24}$$\rm{Fe}_{24}$$\rm{O}_{24}$ are included for every $P$-$T$
condition. Additional compositions
($\rm{Mg}_{36}$$\rm{Fe}_{4}$$\rm{O}_{36}$,
$\rm{Mg}_{30}$$\rm{Fe}_{13}$$\rm{O}_{30}$,
$\rm{Mg}_{16}$$\rm{Fe}_{38}$$\rm{O}_{16}$ and
$\rm{Mg}_{6}\rm{Fe}_{54}\rm{O}_{6}$) were performed for 50 GPa at 5000 and
7500 K, to test the compositional dependence of the Fe-MgO solvus.
Finally, simulations with 1:1 stoichiometries but larger cells
$\rm{Fe}_{64}$, $\rm{Mg}_{30}\rm{Fe}_{30}\rm{O}_{30}$,
$\rm{Mg}_{45}\rm{Fe}_{45}\rm{O}_{45}$ $\rm{Mg}_{40}$$\rm{O}_{40}$, and
$\rm{Mg}_{50}$$\rm{O}_{50}$.

\subsection{Saturation limits} \label{sec:saturation}

$\Delta G_{mix}$ can be related to the volume change associated with the
insertion of an iron of atom into hydrogen, as other contributions are
constant with respect concentration. It can be shown that results for
simulations with a 1:$n$ solute ratio can be generalized to a ratio of
1:$m$ using
\begin{eqnarray}
  \Delta G_c &\approx& F_0(Fe_mMgO)-F_0(Fe_m)-F_0(MgO) \nonumber \\ && -
  \left[F_0(Fe_nMgO)-F_0(Fe_n)-F_0(MgO)\right] \nonumber \\ &=&
  -k_BT\log\left\{ \frac{\left[V(\mathrm{Fe}_n\mathrm{MgO}) +
    \frac{m-n}{n}V(\mathrm{Fe}_n)\right]^{m+1}
    \left[V(\mathrm{Fe}_n)\right]^{n}}
    {\left[V(\mathrm{Fe}_n)\frac{m}{n}\right]^m
    \left[V(\mathrm{Fe}_n\mathrm{MgO})\right]^{n + 1}} \right\},
\end{eqnarray}
where $\Delta G_c = \Delta G_{mix}(1:m)- \Delta G_{mix}(1:n)$, and
$V(\mathrm{Fe}_n)$ and $V(\mathrm{Fe}_n\mathrm{MgO})$ are the volumes for
the simulations of hydrogen and the solution respectively. This allows us
to approximate the saturation limit for MgO in Fe based only on $\Delta
G_{mix}$ and $V$ of our lowest concentration simulation, and $V$ of both of
the endmember compositions.  We note that this low-concentration limit
assumes that the self interaction of the MgO `solute' is negligible in our
lowest concentration, $\rm{Mg}_{6}\rm{Fe}_{54}\rm{O}_{6}$. While this is
not exact, we present it as a estimate for extrapolating these results to
low MgO concentrations. In doing so, we demonstrate that these calculations
are consistent with Mg concentrations the below detection limit of
laser-heated diamond anvil cell experiments performed at lower
temperatures.

\subsection{Comparison of thermodynamic integration with different classical potentials}

\begin{figure}[h!]  
    \centering
    \includegraphics[width=1.0\textwidth]{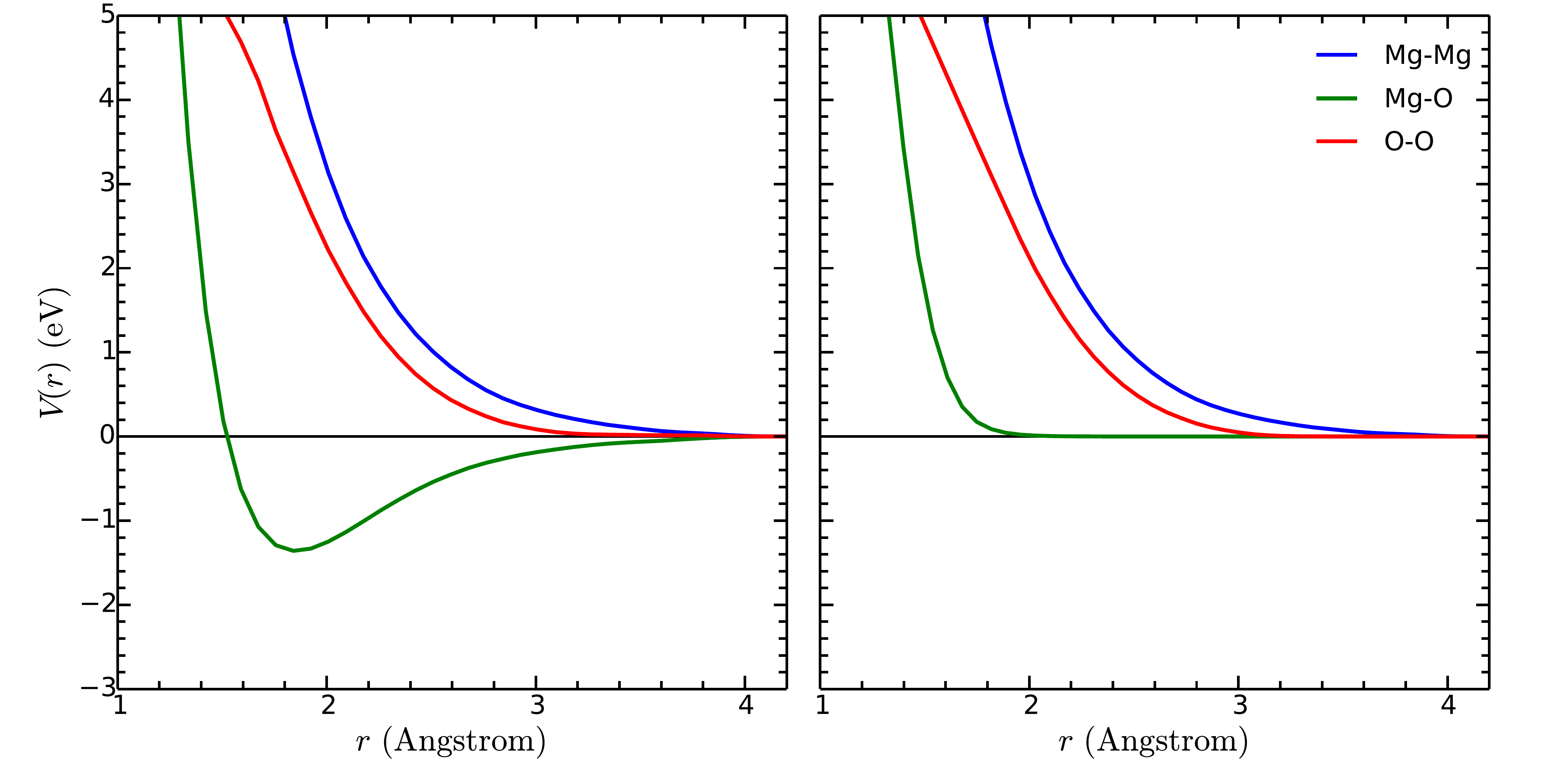}
\caption{Example pair potentials for liquid MgO at 50 GPa and 6000 K. Left: 
  Regular pair potentials fit to DFT-MD simulations, with a linear
  extrapolation at small separation and an asymptote to 0 at large
  separation.  All of the results presented in the paper used this fitting
  procedure. Right: Non-bonding potential fit with the same procedure, but
  constraining values to be positive. Included for comparison with the pair
  potentials in table~\ref{tab:compare_pots}.}
\label{fig:potentials}
\end{figure}

The classical pair potentials are derived by fitting the forces and
positions along a pre-computed DFT-MD trajectory. The potentials are
constructed to approach zero for large separations. For small separations
where the trajectories provide no information, linear extrapolation is
used, which means our pair potentials are finite at the origin. All of the
results presented in the paper used this fitting procedure.
Figure~\ref{fig:potentials} shows an example for the pair potentials for
liquid MgO and 50~GPa and 6000~K.  While the Mg-Mg and O-O potentials are
purely repulsive, the deep mininum in the Mg-O potential represents the
attractive forces between ions of opposite charge.

Table~\ref{tab:compare_pots} provide all terms of the thermodynamic
integration procedure. In order to test how robust our approach is, we
constructed a different set of pair potential where we eliminated all
bonding forces. The values of these non-bonding potentials are constrained
to be positive, asymptoting to zero without a miniumum. Obviously they are a poor representation of the DFT forces
in the system and therefore the free energy differences between the DFT and
the classical system, $F_{\rm cl \to DFT}$, given in table
\ref{tab:compare_pots}, is much larger than for our regular potentials.
However, when the values for $F_{an}$, $F_{an \to cl}$, and $F_{\rm cl \to
DFT}$ are added, we recover the results for $F_{DFT}$ within the 1$\sigma$
error bars. This demonstrates that out free energy calculations are not
sensitive to the details how we construct our classical potentials.

\begin{deluxetable}{ccllllll}
\tabletypesize{\scriptsize}
\tablecolumns{8}
\tablewidth{0pc}
\tablecaption{Comparison of integration paths using different classical potentials. 
  \label{tab:compare_pots}}

\tablehead{ \colhead{Potentials} & \colhead{$F_{\rm an}$} & \colhead{$F_{\rm an \to cl}$} & \colhead{$F_{\rm
 cl}$} & \colhead{$F_{\rm cl \to DFT}$} & \colhead{$F_{\rm DFT}$} }
\startdata
Regular, bonding potentials & $-$427.151 & 27.962 & $-$399.189 & $-$233.825 $\pm$ 0.031 & $-$633.014  $\pm$ 0.031 \\
Non-bonding potentials & $-$427.151 & 204.129 & $-$223.022 & $-$409.927 $\pm$ 0.080 & $-$632.950 $\pm$ 0.080 \\
\enddata
\end{deluxetable}

\begin{figure}[h!]  
  \centering
    \includegraphics[width=0.8\textwidth]{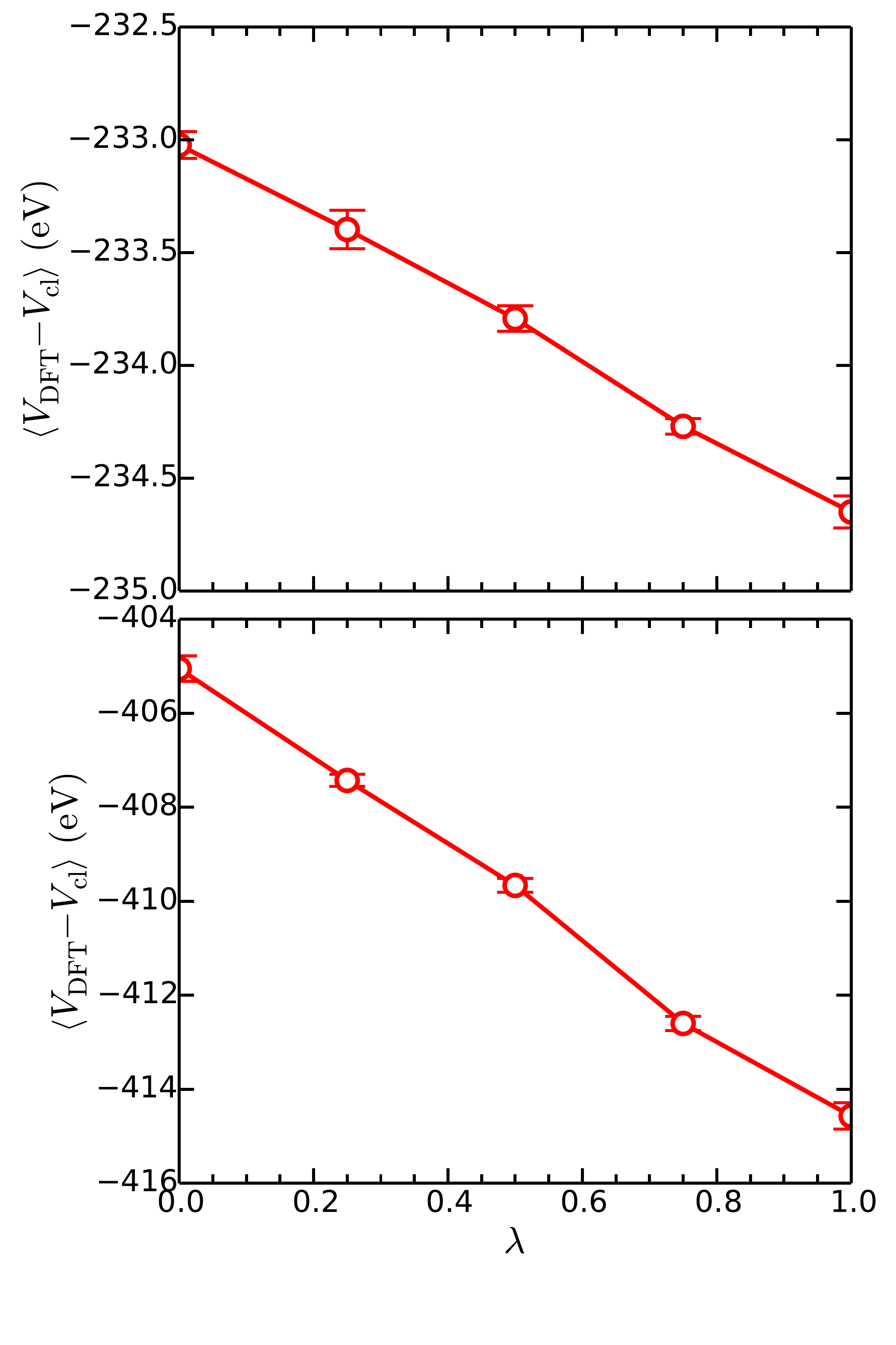}
\caption{The integration path to find $F_{cl \to DFT}$ potentials for MgO
at 50 GPa 6000 K, using the regular, bonding pair potentials (upper) and
the non-bonding pair potentials (lower). }
\label{fig:integrate_dft}
\end{figure}

Although the correct final result is found when using an unrealistic
potential, the efficiency for the thermodynamic integration is the highest
when the classical forces best match the DFT forces.
Figure~\ref{fig:integrate_dft} shows the calculated values of $\left<
V_{DFT} - V_{\rm cl} \right>$ as a function of $\lambda$ using the regular
``bonding'' potential, and for the ``non-bonding potential''. Here each
plotted value of lambda represents and independent DFT-md simulation with
using that fraction of DFT forces, along with the complementary fraction of
classical forces. The integral of this function give the helmholtz free
energy $F_{\rm cl \to DFT}$. In the first case , the function $\left<
V_{DFT}-V_{\rm cl}\right>$ in figure~\ref{fig:integrate_dft} depends weakly
on $\lambda$. The function is almost linear and the differnce between
values at $\lambda=0$ and 1 is small. When both criteria are satisfied,
very few $\lambda$ points are needed to evaluate the integral. The
simulations with non-bonding potentials experience larger fluctuations  due
to the greater mismatch in forces, leading to a larger statistical
uncertainty.  As a result, these simulations required longer simulation
times to match the results found with bonding potentials.

\begin{figure}[h!]  
  \centering
  \includegraphics[width=1.0\textwidth]{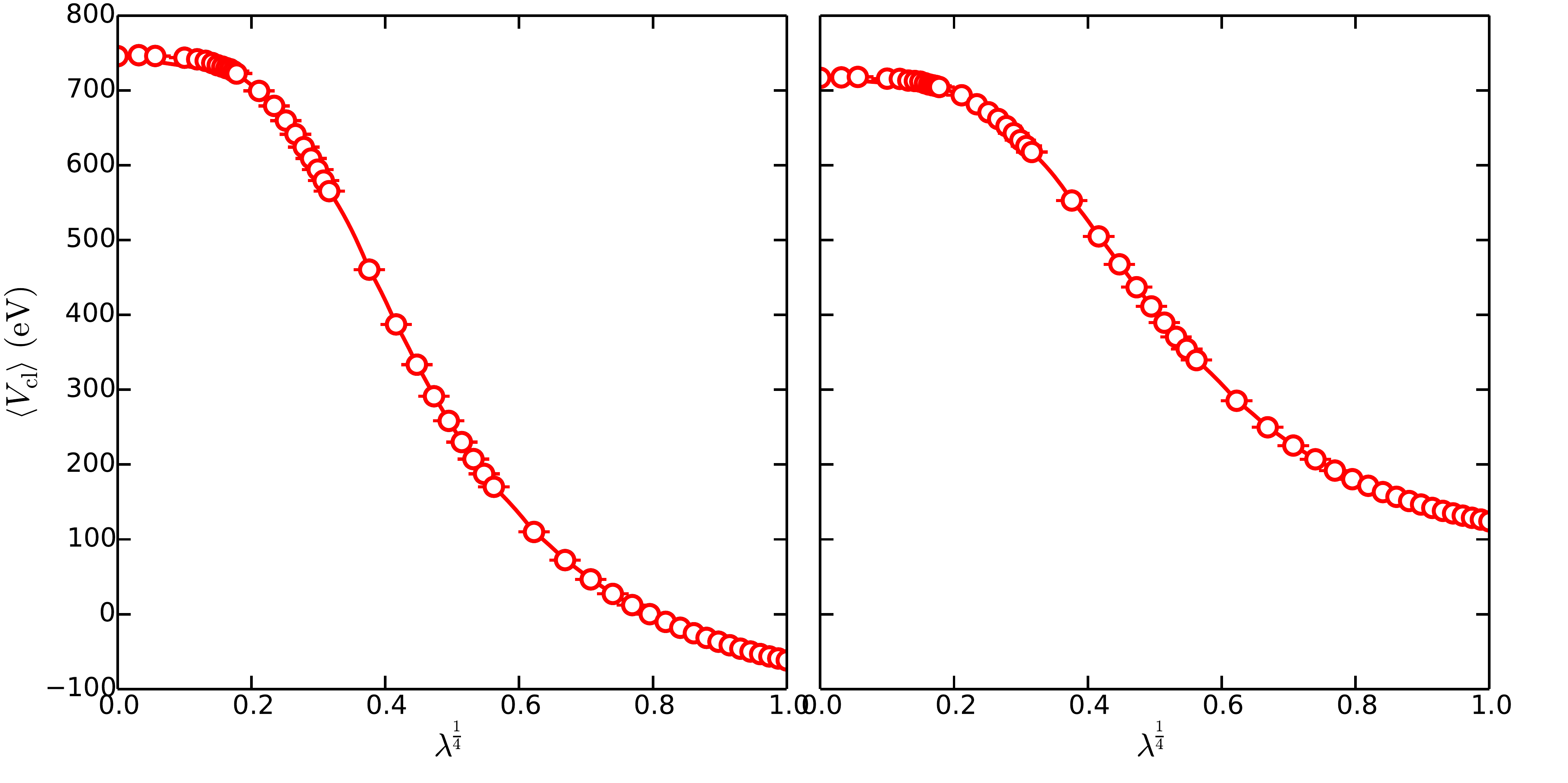}
\caption{The integration path to find $F_{an \to cl}$ potentials for MgO at 50 GPa 6000
K, using the regular, bonding pair potentials (left) and
the non-bonding pair potentials (right). These are plotted against the
integration parameter, $\lambda$, to the $1/4$ power.}
\label{fig:integrate_cmc}
\end{figure}

Figure~\ref{fig:integrate_cmc} shows $\left< V_{\rm cl} \right>$ as a function
of $\lambda$ from classical monte carlo simulations. The integration of
$F_{an \to cl}$function becomes strongly non-linear as $\lambda$
approaches zero, the non-interacting case. Since classical simulations are
approximately 10$^5$ times more efficient, it is possible to obtain very
close sampling of the a cusp in the integrated function. 

Using the definition for the ensemble averaged potential at a given
$\lambda$
\begin{eqnarray}
  \left< V_{\rm cl} \right>_{\lambda} = \frac{ \int \! d{\bf r} \, V_{\rm cl}({\bf r}) 
  e^{ -\beta \lambda V_{\rm cl}({\bf r}) } }
  { Z}, \label{eqn:average}
\end{eqnarray}
where $Z$ is the partition function
\begin{eqnarray}
Z =  \int \! d{\bf r} \, e^{ -\beta \lambda V_{\rm cl}({\bf r}) },
\end{eqnarray}
we find the following expression as $\lambda \to 0$
\begin{eqnarray}
 \left< V_{\rm cl} \right>_{\lambda \to 0} &=& \frac{ \int \! d{\bf r} \, V_{\rm
 cl}}
 {  \int \! d{\bf r} \, 1} \nonumber \\
 &=& \frac{1}{V}  \int \! dr \, r^2 V_{\rm cl}(r) \label{eqn:limit}
\end{eqnarray}
Then from the deriviative of $\left< V_{\rm cl} \right>_{\lambda}$ in
equation \ref{eqn:average} and \ref{eqn:limit}
\begin{eqnarray}
  \frac{d \left< V_{\rm cl} \right>}{d \lambda}
   &=& \frac{1}{Z^2} 
  \left[ Z \int \! d{\bf r} \, \left(-\beta \right)V_{\rm cl}^2({\bf r}) 
    e^{ -\beta \lambda V_{\rm cl}({\bf r}) } 
   - \left(-\beta \right)\left\{  \int \! d{\bf r} \, V_{\rm cl}({\bf r}) e^{ -\beta \lambda
     V_{\rm cl}({\bf r}) } \right\}^2
    \right] \nonumber \\
   &=& (-\beta) \left[ \left< V_{\rm cl}^2 \right> - \left< V_{\rm
   cl}\right>^2 \right] \nonumber \\
   \left. \frac{d \left< V_{\rm cl} \right>}{d \lambda}\right|_{\lambda \to 0} 
   &=& (-\beta) \left[  \frac{1}{V}  \int \! dr \, r^2 V^2_{\rm cl}(r) -
     \left\{ \frac{1}{V}  \int \! dr \, r^2 V_{\rm cl}(r) \right\}^2 \right]
  \end{eqnarray}
This give us the slope and intercept for the integration at $\lambda=0$,
necessary to correctly integrate the cusp.  Becauase of the extreme
difference in computational efficiency, it is always best to adjust the
classical potential to match the DFT forces.

\subsection{Verification of thermodynamic integration in multicomponent
systems}
The second test is to verify that, in a multi-component system, the
integration path does not effect the results. An integration path needs to
be constructed that connects a system with Mg-Mg, Mg-O, and O-O pair
potentials with an non-interacting system. In our standard proceduce we
turn on all pair potentials simultaneously by changing
$\lambda_1=\lambda_2=\lambda_3$ from 0 to 1.
\begin{equation}
  V_{\lambda_1\lambda_2\lambda_3} = \lambda_1 V_{\rm Mg-Mg} + \lambda_2 V_{\rm Mg-O} + \lambda_3 V_{\rm O-O} 
\end{equation}
However, as we will now demonstrate, alternative integration paths will give the same results. We compare different integration paths to calculate
the free energy of the classical system, $F_{\rm cl}$, for both the regular
and non-bonding potentials. This is comparison can not be made directly for
$F_{\rm cl \to DFT}$ because tracking the intereaction of different species
separarately is not possible in a Kohn-Sham formulation.

\begin{deluxetable}{clllllll}
\tabletypesize{\scriptsize}
\tablecolumns{8}
\tablewidth{0pc}
\tablecaption{Comparison of different integration paths using classical potentials. 
  \label{tab:compare_int_path}}

\tablehead{ \colhead{Potential} & 
            \colhead{$F_{\rm an}$} & 
            \colhead{$F_{\rm step}$} & 
            \colhead{$F_{\rm an} + \sum F_{\rm step}$}}
\startdata
Non-bonding & $-$427.151 & $F_{000 \to 111}$ = 204.196(27)      & $-$222.955(27)
\\[1mm]\hline\\[-2.5mm]
Non-bonding & $-$427.151 & $F_{000 \to 010}$ = $\,\:$16.328(6)  & $-$222.953(45)\\
            &          & $F_{010 \to 111}$ = 187.870(39) 
\\[1mm]\hline\\[-2.5mm]
Non-bonding & $-$427.151 & $F_{000 \to 101}$ = 166.363(15)      & $-$222.962(21)\\
            &          & $F_{101 \to 111}$ = $\,\:$37.826(18)
\\[1mm]\hline\\[-2.5mm]
Non-bonding & $-$427.151 & $F_{000 \to 100}$ = 104.323(11)      & $-$222.954(52)\\
            &          & $F_{100 \to 110}$ = $\,\:$31.225(10) \\
            &          & $F_{110 \to 111}$ = $\,\:$68.650(31)
\\[1mm]\hline\\[-2.5mm]
Regular     & $-$427.151 & $F_{000 \to 111}$ = $\,\:$28.028(31) & $-$399.123(31) 
\\[1mm]\hline\\[-2.5mm]
Regular     & $-$427.151 & $F_{000 \to 101}$ = ~~182.820(16)      & $-$399.123(37) \\
            &          & $F_{101 \to 111}$ = $-$154.792(21)
\enddata
\end{deluxetable}

In line 2 of table~\ref{tab:compare_int_path}, we turn on the Mg-O
potential in the first integration step ($F_{000 \to 010}$) and then switch
on the Mg-Mg and O-O potentials in the second and final integration step
($F_{101 \to 111}$). The indices refer to the three $\lambda$ values for
Mg-Mg, Mg-O, and O-O potentials, respectively. In line 3 of
table~\ref{tab:compare_int_path}, we interchange both integration steps. In
line 4, we performed three integration steps turning on the one potential
after the other. In the last column we compare the classical free energies
after adding the results from every integration step to $F_{\rm an}$. The
results agree within the statistical uncertainties demonstrating that the
same classical free energies can be obtained for four different integration
paths using non-bonding potentials.

In table~\ref{tab:compare_int_path}, we also show the results for two
integration paths using regular, bonding potentials. We find consistent
results when we either turn on all potentials simultaneously and when we
switch on the Mg-Mg and O-O potentials in the first step and the Mg-O
potential in the second. We were not able, however, to turn on the
attractive Mg-O alone because the system becomes unstable due to the
imbalance between attractive and missing repulsive forces. This is similar
to what happens in the case of a first-order phase transition, which over
which thermodynamic integrations are also invalid. Nevertheless, this test
demonstrates that different integration paths give consistent results also
for the systems with attractive forces when care is taken to taken to avoid
instabilities.




